\documentclass[a4paper,fleqn,usenatbib]{mnras}

\usepackage[T1]{fontenc}
\usepackage{ae,aecompl}

\usepackage{graphicx}	
\usepackage{amsmath}	
\usepackage{amssymb}	

\newcommand{\press}{{\ \rm cm}^{-3}({\rm km/s})^2}
\newcommand{\msunsigma}{{\ \rm M}_\odot/{\rm pc}^2}
\newcommand{\vsigma}{{\ \rm km/s \, M}_\odot/{\rm pc}^2}
\newcommand{\speed}{{\ \rm km/s}}
\newcommand{\density}{{\ \rm cm}^{-3}}

\title[RPS made easy]{Ram Pressure Stripping Made Easy: An Analytical Approach}

\author[J. K\"oppen et al.]{
J. K\"oppen,$^{1,2,3}$\thanks{E-mail: koeppen@astrophysik.uni-kiel.de}
P. J\'achym,$^{1}$
R. Taylor,$^{1}$
and J. Palou\v s $^{1}$
\\
$^{1}$Astronomical Institute,
      Academy of Sciences of the Czech Republic,
      Bo\v cn\' \i \ II 1401, 141 31 Prague 4, Czech Republic\\
$^{2}$Observatoire Astronomique de Strasbourg, Universit\'e de Strasbourg,
      CNRS, UMR 7750, 11 Rue de l'Universit\'e,
      F-67000 Strasbourg, France\\
$^{3}$Institut f\"ur Theoretische Physik und Astrophysik,
      Universit\"at Kiel,
      D--24098 Kiel, Germany\\
}

\date{Accepted 2018 June14. Received 2018 June 14; in original form 2017 November 28}

\pubyear{2018}

\begin{document}
\label{firstpage}
\pagerange{\pageref{firstpage}--\pageref{lastpage}}
\maketitle

\begin{abstract}
  The removal of gas by ram pressure stripping of galaxies is treated by a purely 
  kinematic description. The solution has two asymptotic limits: if the duration of
  the ram pressure pulse exceeds the period of vertical oscillations perpendicular
  to the galactic plane, the commonly used quasi-static criterion of Gunn \& Gott 
  is obtained   which uses the maximum ram pressure that the galaxy has experienced 
  along its orbit. For shorter pulses the outcome depends on the time-integrated 
  ram pressure. This parameter pair fully describes the gas mass fraction that is 
  stripped from a given galaxy. This approach closely reproduces results from SPH 
  simulations.   We show that typical galaxies follow a very tight relation in this 
  parameter space corresponding to a pressure pulse length of about 300 Myr. 
  Thus, the Gunn \& Gott criterion provides a good description for galaxies in larger 
  clusters. Applying the analytic description to a sample of 232 Virgo galaxies 
  from the GoldMine database, we show that the ICM provides indeed the ram pressures 
  needed to explain the deficiencies. We also can distinguish current and past strippers, 
  including objects whose stripping state was unknown. 
\end{abstract}




\section{Introduction}

  In clusters of galaxies one finds spiral galaxies with a quite normal appearence in the 
  optical region, but whose HI mass is significantly less than expected for a normal spiral 
  of the same type and size. This HI deficiency can be understood as the result of ram 
  pressure stripping from the interaction of the galaxy's interstellar matter (ISM) 
  with the hot intra cluster medium (ICM) that fills a cluster and is detected by its X-ray 
  emission, and through which the galaxy travels during its flight in the cluster. 
  
  In their study of infall of matter into the centers of clusters of galaxies, \citet{gunngott72} 
  estimated that the gas would be removed from a spiral galaxy if the ram 
  pressure exceeded the force density from the galactic disk's gravitational field. 
  Since then ram pressure stripping has been investigated by several numerical techniques 
  (see \citet{schulz01} for an overview of the early work). The SPH simulations by 
  \citet{abadi99} showed that the stripping radius, i.e. the outer radius of the remnant 
  HI disk, is reasonably well predicted analytically, based on the ideas of \citet{gunngott72}
  using the gravitational attraction of bulge, dark halo, and an infinitesimally thin disk.
  \citet{vollmer01} found that the results of his sticky-particle models could well be matched 
  with the predictions of \citet{gunngott72} using the assumption that the maximum restoring 
  force from the potential of bulge, dark halo, and a disk with finite thickness
  equals the centrifugal force at the stripping radius. Also, fitting formulae were 
  obtained for the fractions of the stripped mass and the reaccreted mass, as a function 
  of maximum ram pressure and tilt angle. The comprehensive 2-D hydrodynamical simulations 
  of galaxies subjected to a constant flow of hot ICM by \cite{roediger05} showed that the 
  stripping radius and stripped mass fractions are a function of ram pressure, but not 
  of ICM density or speed separately. The numerical results could closely be reproduced with 
  the \cite{gunngott72} approach by computing the maximum restoring force.

  The SPH simulations by \cite{jachym07} cover a wide range of ICM density and spatial 
  extent of the ICM showing how the amount of stripped mass depends on the duration of 
  the pressure pulse. It is found that while the runs with longest pulse durations could be 
  matched by the predictions using the \citet{gunngott72} approach, at short pulses stripping
  relies on the amount of momentum transferred to the galaxy's gas. 
  The tilt angle of the ICM wind plays only a minor role, as long as the angle between the
  ICM flow direction and normal to the plane of the gas disk is smaller than about $60^o$, i.e.
  when stripping occurs nearly face-on \citep{roediger06, jachym09}. 
  The high-resolution three-dimensional hydrodynamical simulations of stripping of
  galaxies with a clumpy, multiphase ISM by \citet{tonnesen09} indicate that low 
  density gas at any position in the galaxy is quickly removed. High-density 
  clouds are subject to ablation and later are eventually stripped. While it is difficult
  to quantify how much gas is lost from the disk area inside the stripping radius, 
  this work demonstrates that gas loss from the inner disk might be substantial.    

  The sticky-particle approach \citep{vollmer01} enabled detailed 
  analyses of a number of HI deficient galaxies in the Virgo cluster which lead 
  \citet{vollmer09} to put them in a sequence beginning from galaxies which just start to 
  be affected by stripping to objects that completed the stripping process and are already 
  past the pressure maximum. In these studies the observational data must be detailed, such 
  as having HI maps and data cubes in order to match the density and velocity structure of 
  the HI flowing away from the host galaxy. Due to the large computational effort this 
  technique could only be applied for a limited number of the more prominent galaxies.

  It would be advantageous to have more efficient techniques requiring less computational 
  modelling that would allow a reliable interpretation based on less detailed data which are
  available for a larger number of objects. As comparison of SPH simulations 
  \citep{jachym07, jachym09} with analytical considerations showed that the numerical results 
  could be well understood in terms of a substantially simpler kinematical approach. The 
  present paper addresses a more complete formulation describing ram pressure stripping 
  which is done by considering the motions of test particles approximating gas parcels 
  of a galaxy in the galaxy's gravitational potential and subjected to an external force pulse. 

  This analysis allows us to give a comprehensive description how strongly a galaxy is 
  stripped for a given ram pressure history (Section 2). With this analytics we assess 
  the sensitivity of the results on the various parameters of the galaxy (Section 3). 
  This also imposes constraints on the ram pressure histories under which galaxies are 
  stripped (Section 4). Section 5 shows that for the most likely trajectories of 
  a galaxy falling into a cluster centre, the ram pressure histories are well constrained, 
  and that for observed clusters the long-pulse limit is an appropriate approximation.
  In Section 6 we compare the analytical approach with SPH simulations \citep{jachym07, 
  jachym09} and explore the mass fractions of the stripped gas and the gas which will
  fall back onto the disk. In Section 7 we describe our method how to interpret global data 
  from galaxies and apply it to data from the GoldMine database to identify galaxies 
  which are either currently undergoing stripping or having had a stripping event in the past.

  Several interactive JavaScript tools which evaluate the analytical considerations of
  this paper are available in the internet
  at http://www.astrophysik.uni-kiel.de/$\sim$koeppen/JS/RPShome.html.

\section{Motion of a gas parcel in a galaxy}

\begin{figure}
  \includegraphics[width=0.33\textwidth, angle=270]{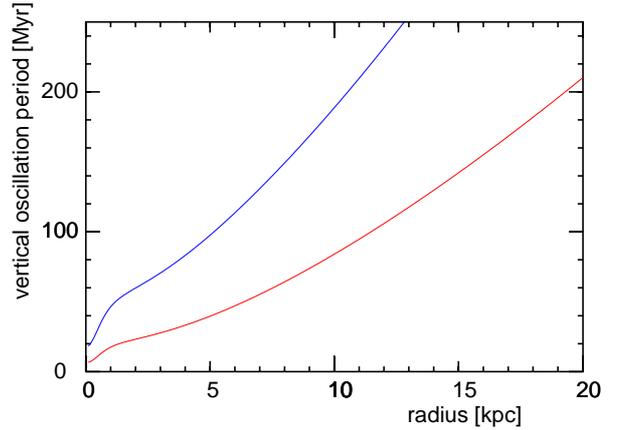}
  \caption{
    Variation of the period of vertical oscillations about the galactic plane as a function 
    of distance from the centre of a galaxy like the Milky Way (red curve). The blue curve 
    is for NGC 4522 as an example for a smaller galaxy.
    }
  \label{f:vertperiod}
\end{figure}

  Let us consider the motion of a parcel of gas in the galactic disk like a single mass point,
  in the gravitational field of the galaxy and under the influence of an external pressure pulse. 
  Thereby we neglect any internal gas dynamics of the disk and replacing the hydrodynamic 
  ISM/ICM interaction by a pressure pulse. In the plane of the disk this parcel will be at the 
  minimum of the gravitational potential at its distance from the rotational axis. If slightly 
  perturbed in a direction perpendicular to the plane, it will start vertical oscillations of 
  small amplitude. The period of these oscillations depends on the vertical derivative of the 
  force in the galactic plane
\begin{equation} 
   T_{\rm vert} = 2\pi/\sqrt{\partial^2 \Phi(r,z)/\partial z^2|_{z=0}}
\end{equation} 
  For a galaxy similar to the Milky Way it will vary from around 10~Myr near the centre to 
  200~Myr near the outer edge, as shown in Fig.\ref{f:vertperiod}.

  During a stripping event, ram pressure produces an additional force acting on the gas parcel 
  (with surface density $\Sigma_g$). In the case of face-on stripping, this external force is 
  only in vertical direction with respect to the galactic plane (the x-y plane):
\begin{eqnarray} 
   \Sigma_g {d v_x\over dt} &=& -  \Sigma_g {\partial \Phi \over \partial x} \label{e:eomx}\\  
   \Sigma_g {d v_z\over dt} &=& -  \Sigma_g {\partial \Phi \over \partial y} \label{e:eomy}\\  
   \Sigma_g {d v_z\over dt} &=& p(t) -  \Sigma_g {\partial \Phi \over \partial z \label{e:eomz}}  
\end{eqnarray}
  where Eqns.\ref{e:eomx} and \ref{e:eomy} define the components in the galaxy's rotational plane
  and Eq.\ref{e:eomz} gives its component perpendicular to that plane.  
  We assume the surface density $\Sigma_g$ of the parcel is specified for a given 
  galactocentric distance $R$ and is constant, i.e. it does not depend on the height $z$ above
  the galactic plane. The time-varying force $p(t)$ is given by the ram pressure
\begin{equation} 
   p(t) = \rho_{\rm ICM} (v_{\rm ICM} - v_z)^2
\end{equation} 
  where $\rho_{\rm ICM}$ is the density of the ICM at a place on the galaxy's
  orbit in the cluster, $v_{\rm ICM}$ is the velocity of the galaxy relative to the ICM and
  $v_z$ is the velocity of the parcel of gas relative to the parent
  galaxy. $v_{\rm ICM}$ is of the order of 1000 km/s, thus well above $v_z$ which is 
  mainly due to the vertical gas motions inside the galaxy (about 10 km/s) and later 
  eventually increased by the action of ram pressure forces. This means that we may 
  disregard the contribution to the ram pressure due to the galaxy's internal motions.

\subsection{Long-pulse limit: The Gunn \& Gott criterion}

  \citet{gunngott72} pointed out that gas is held back in the disk of the galaxy, as long as 
  the ram pressure does not exceed the restoring force by the gravitational potential. In the 
  potential of a disk galaxy at some galactocentric distance the restoring force 
  ${\partial \Phi\over \partial z}(z)$ has a maximum value at some height $z$ above the disk. 
  Taking this as a threshold, the condition of stripping gas from that radius is formulated as:
\begin{equation} \label{e:longpulse}
   p_{\rm max} \ge \Sigma_g(r) \bigg|{\partial \Phi(r,z) \over \partial z}\bigg|_{\rm max} 
\end{equation} 
  with the maximum ram pressure $p_{\rm max}$ that has occurred so far on the galaxy's
  orbit in the cluster. The height above the plane at which the maximum restoring force occurs 
  can be obtained from ${\partial^2 \Phi /\partial z^2} = 0$. In this quasi-static approach 
  one assumes that the ram pressure increases slowly enough so that the external force remains 
  always balanced by the restoring force, and that it lasts until the gas element reaches the 
  height of maximum restoring force. But even if the ram pressure overcomes this maximum, the 
  gas element still remains gravitationally bound to the galaxy. The pressure has to continue 
  until the total energy of the gas element becomes positive and it can escape with its
  random velocity. This is explored in more detail in Appendix \ref{s:long}.

  The time it takes the gas element to reach the height of maximum restoring force is 
  of the order of the period of vertical oscillations.  Thus, Eq.\ref{e:longpulse} 
  forms a criterion for the removal of the gas if the duration of the ram pressure 
  pulse is longer than the period of vertical oscillations. Tests with models for spiral 
  galaxies show that the escape speed at the height of maximum restoring force is still 
  more than 90\% of the escape speed in the galactic plane. Thus, the long duration of
  a force exceeding\footnote{Since the local value of the restoring force 
  decreases beyond the height of this maximum, the ram pressure force needs to be
  only comparable or can later be even less than the maximum value in order to provide the
  work necessary for escape.} the maximum restoring force is an essential ingredient 
  of the criterion from \citet{gunngott72}. We shall name this the long-pulse limit.

  As the maximum restoring force decreases outwards in a galactic disk, the gas is lost 
  outside some radius $r$ where the equality sign holds in Eq.\ref{e:longpulse}. Thus the gas 
  disk is truncated at this radius which might be called the {\it stripping radius}. 
  On its orbit towards the interior of a cluster with its higher ICM density, 
  a galaxy experiences an increasing ram pressure, and the galaxy's gas disk --
  whose surface density usually decreases outward -- is progressively truncated 
  until some minimum galactocentric radius is reached. As a consequence, the outcome of the stripping 
  event, the gas mass fraction remaining in the disk, depends on the maximum ram 
  pressure that the galaxy had so far been subjected to.

\subsection{Short-pulse limit: Momentum transfer}

  At the other extreme end, let us consider a force pulse much shorter than the  period for vertical 
  oscillations: As in the classical apparatus of the ballistic galvanometer we may consider the parcel 
  staying at rest in the galactic plane during the duration of the pulse. The parcel of gas with 
  surface density $\Sigma_g$ will get all the momentum provided by the pulse, and at the end of the 
  pulse it will have accumulated the momentum
\begin{equation} 
   \Sigma_g  v_{\rm AFTER} = \int p(t) dt
\end{equation} 
  where $v_{\rm AFTER}$ is the gas parcel's velocity after the pulse. The accumulated momentum is 
  nothing but the ram pressure integrated over the pulse duration. We add the momentum but not the 
  mass of the intracluster medium; the original mass and surface density of the accelerated gas 
  parcel remain constant. This is justified by the fact that the momentum provided by the ICM is 
  connected to the large velocity of the galaxy relative to the ICM, but the ICM density is much 
  smaller than the ISM, thus we may disregard it.

  Subsequently, the parcel would perform vertical oscillations whose amplitude depends on 
  $v_{\rm AFTER}$. If this speed exceeds the local escape speed $v^2_{\rm esc} = -2\Phi$, 
  corrected for the kinetic energy of the parcel's orbital motion
\begin{equation}
   v_{\rm AFTER} \ge \sqrt{-2\Phi -v^2_{\rm circ}} 
\end{equation} 
  the parcel will get stripped.  This means that in the short-pulse limit the outcome of a stripping 
  event is determined by the time integral over the ram pressure rather than its maximum value: 
\begin{equation} \label{e:shortpulse}
   \int p(t) dt \ge \Sigma_g \sqrt{-2\Phi -v^2_{\rm circ}} 
\end{equation} 
  This is quite different from the long-pulse limit. Furthermore, the outcome of a stripping event is 
  independent of the shape of the pressure pulse. The momentum transfer for longer pulses is addressed 
  in more detail in Appendix \ref{s:efficiency}. The time-integrated force 
\begin{eqnarray} 
   \int p(t) dt & = & \int \rho_{\rm ICM} v^2_{\rm ICM} dt \equiv (v\Sigma)_{\rm ICM} \\
                & \approx & 
                  \hat{v}_{\rm ICM} \int \rho_{\rm ICM} v_{\rm ICM} dt    
     = \hat{v}_{\rm ICM} \Sigma_{\rm ICM} \nonumber
\end{eqnarray} 
  is related to the column density $\Sigma_{\rm ICM}$ intercepted by the parcel during the galaxy's 
  flight through the ICM, as introduced by \citet{jachym07}, with the speed $\hat{v}_{\rm ICM}$ 
  of the galaxy near the time of maximum pressure.  
  For convenience, we shall give $(v\Sigma)_{\rm ICM}$ in units of $1000\vsigma$. 

\subsection{The Combined Diagram}\label{s:combine}

  From the two limiting cases it is apparent that there are two parameters, $v\Sigma_{\rm ICM}$ 
  and $p_{\rm max}$, which determine the outcome of a stripping event for a given galaxy. Thus, 
  the diagram of the plane spanned by these two parameters is appropriate for representing the 
  solutions of ram pressure stripping. 

  A typical spiral galaxy is composed of the stellar bulge, the stellar disk, the gas disk,
  and the dark matter halo, which form the gravitational potential. Each of the components
  is described by parameters, e.g. as in Table \ref{t:LMparms} for models that we 
  use in this paper. Given these properties of a galaxy, the stripped mass fractions can be 
  estimated for any pair ($v\Sigma_{\rm ICM}$, $p_{\rm max}$): Equations \ref{e:longpulse} 
  and \ref{e:shortpulse} allow us to predict above which radius the gas is removed from the 
  galactic disk. Integration of the gas mass outside this {\it stripping radius} then gives 
  the stripped mass fraction. 

\begin{figure}
  \includegraphics[width=0.33\textwidth, angle=270]{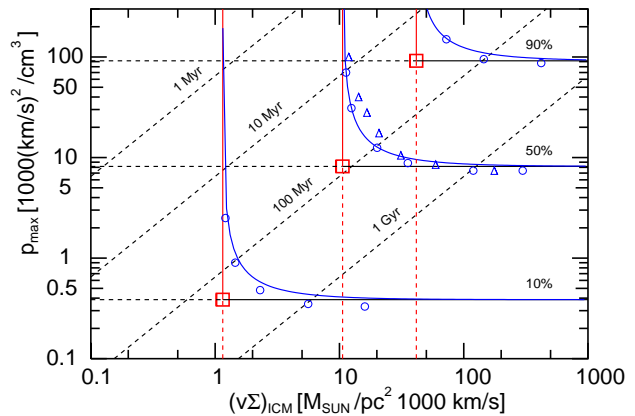}
  \caption{
    Loci in the plane of maximum $p_{\rm max}$ and time-integrated ram pressure 
    $(v\Sigma)_{\rm ICM}$ predicted for 10, 50, and 90\% stripped mass fractions in 
    a Milky Way type galaxy (model LM4 with parameters given in Table \ref{t:LMparms}): 
    The horizontal black line is the long-pulse limit \citep{gunngott72}, the vertical 
    red line marks the short-pulse limit. Both are joined with an interpolating curve 
    (in blue). Blue open circles and triangles are results from test particle models that 
    yield the respective stripped mass fractions with Gaussian and Lorentz pulses,
    respectively. Red squares mark the crossing points of long- 
    and short-pulse limits. Black dashed diagonal lines give the $p_{\rm max}$ versus 
    $(v\Sigma)_{\rm ICM}$ relation for different durations for ram pressure pulse with
    Gaussian shape.}
  \label{f:theplane}
\end{figure}

  In Fig.\ref{f:theplane} we show the predictions for a Milky Way type galaxy (model LM4, 
  with parameters given in Table \ref{t:LMparms}) for stripped mass fractions of 
  10, 50, and 90\%. At 10\% stripped fraction only gas from the outer rim of the galaxy 
  is removed, while at 90\% the galaxy is heavily stripped down to its inner parts.
  The horizontal line is the long-pulse limit \citep{gunngott72}, the vertical line 
  marks the short-pulse limit. As a given galaxy can be stripped to the same given level 
  by different combinations of $p_{\rm max}$ and $v\Sigma_{\rm ICM}$ values, i.e. by 
  pulses of different duration and strength -- from short/strong to long/weak --
  the complete geometrical locus for a given stripped fraction is represented by a curve 
  which joins the two limiting cases. A rather simple interpolation is suitable
  \begin{equation} \label{e:transit}
     p_{\rm max} = p_0 (1 + 0.5/((v\Sigma)_{\rm ICM}/(v\Sigma)_0 -1)) 
  \end{equation} 
  to provide a smooth transition between the slopes of the two limiting laws.  
  ($(v\Sigma)_0$, $p_0$) are the coordinates of the crossing point of short- and long-pulse 
  limits which specifies the minimum pulse duration and the minimum peak ram pressure 
  that would be necessary to strip the galaxy to a given stripped mass fraction.

  The figure also shows that for a Milky Way type galaxy the criterion of 
  \citet{gunngott72} predicts the maximum ram pressure necessary for a certain stripping 
  outcome only if the pulse duration is longer than about 200 Myr for the outer rim 
  (see Fig.\ref{f:vertperiod}) and less for the inner parts, i.e. if the time integrated 
  pressure exceeds a certain value. Shorter pulses require correspondingly higher maximum 
  ram pressures.

  Conversely, one may predict which conditions ($v\Sigma_{\rm ICM}$, $p_{\rm max}$) are 
  necessary for a given galaxy to lose a given fraction of its gas. As these calculations 
  depend on the galaxy's parameters, the position of a stripping event in this diagram 
  gives some information about the galaxy. Regions in the diagram can be identified 
  in which normal galaxies are expected (Sect.\ref{s:others}).

  The diagram has a second aspect: For a given shape of the ram pressure pulse its 
  time-averaged value and its maximum value completely describe the history of the event. 
  As this history is the result of the galaxy's flight trajectory through the cluster and 
  the distribution of the ICM therein, this diagram can also serve to place constraints on 
  the galaxy's orbit. In Sect.\ref{s:orbits} we identify the regions for most probable 
  orbits in clusters of galaxies.

\subsubsection{Test Particle Models} \label{s:ring}

  To allow computation of intermediate cases between the two limits, and to facilitate   
  an efficient exploration of how the outcome of stripping events depends on the 
  numerous parameters of the galaxy and the ram pressure history, it is convenient to 
  use numerical models, based on the purely kinematical description of the movement 
  of gas parcels, without any hydrodynamical effects. We consider the evolution of 
  the gas disk in a typical spiral galaxy which is subjected to face-on ram pressure. 
  The time-dependence of the ram pressure is modelled as a Gaussian or Lorentzian 
  shaped pulse of specified maximum value $p_{\rm max}$ and duration (FWHM). The galaxy
  is composed of bulge, stellar and gas disk, and dark halo, which determine the
  gravitational potential. As an example, we use the Milky-Way type model LM4 with the 
  parameters given in Table \ref{t:LMparms}. The gas disk is represented by a number 
  (usually 1000) of test particles, each standing for an entire mass ring in the galactic 
  disk. The response to the ram pressure force is followed for each particle by solving 
  its equation of motion in three dimensions (Eqs.\ref{e:eomx} - \ref{e:eomz}) due to 
  the galaxy's gravitational potential and the ram pressure pulse only. There is no 
  interaction between individual particles. Each particle is given a mass proportional to 
  the surface density at its initial galactocentric radius, and is placed in a circular 
  orbit following the rotation curve. The equations of motion are integrated with a 4th order 
  symplectic Runge-Kutta type integrator \citep{forest90} and adaptive time stepping. 

  The results match the short-pulse limit quite accurately (Fig.\ref{f:theplane}), but 
  they require a slightly lower maximum pressure than the long-pulse limit. This is partly 
  because with a constant force the minimum force to dislodge gas parcels is slightly 
  less than the quasi-static estimate, but also because during the long time of the 
  interaction a test particle is pushed by the external pulse from the initial circular 
  orbit into a box orbit, in which it performs oscillations in vertical but also in radial 
  direction. From a more outward position it is easier to escape. Apart from this small 
  effect, the test particle models demonstrate the validity of the simple interpolating 
  curve between long and short-pulse limits for both Gaussian and Lorentzian shaped pulses. 

\section{Sensitivity of Parameters}

\begin{figure}
  \includegraphics[width=0.33\textwidth, angle=270]{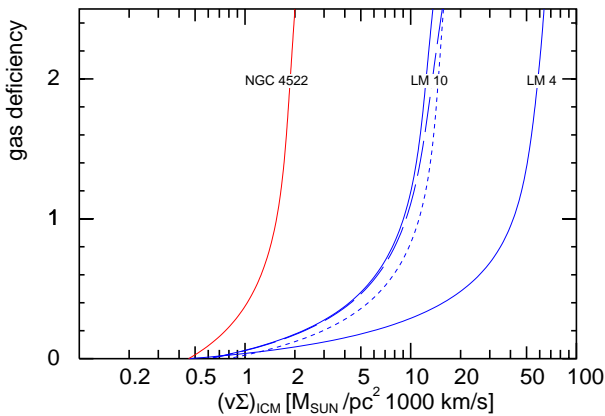}
  \caption{
    The gas deficiency as a function of the time integrated ram pressure computed 
    for the short-pulse limit for the Milky Way-like LM4 and LM10 model galaxies 
    and for NGC~4522 \citep{vollmer00}. Broken lines refer to the LM10 model with 
    either the bulge (long dashes) or the dark halo (short dashes) mass doubled.}
  \label{f:defshort}
\end{figure}

\begin{figure}
  \includegraphics[width=0.33\textwidth, angle=270]{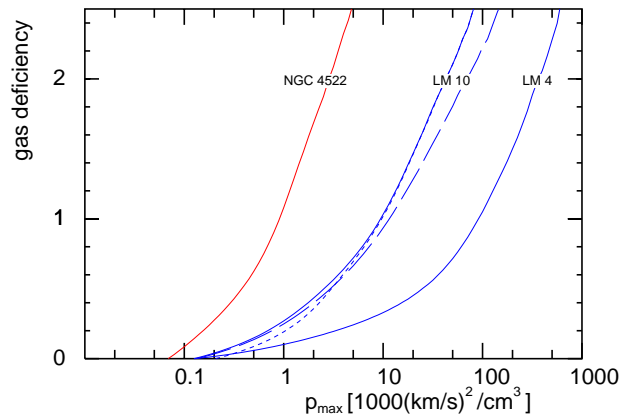}
  \caption{
    Like Fig.\ref{f:defshort}, but for the long-pulse limit.}
  \label{f:deflong}
\end{figure}

  The analytical approach permits us to evaluate more easily the dependence of the 
  expected stripped mass fractions by short- and long-pulses on the numerous parameters 
  of a model galaxy. For some typical examples we show in Figs.\ref{f:defshort} and 
  \ref{f:deflong} the results for the short- and long-pulse limits. The deficiency 
  $def = \log_{\rm 10}(M_0/M) = -\log_{\rm 10}(1 - m_{\rm strip})$ is computed from 
  the gas mass $M_0$ before and after ($M$) the stripping event, or the stripped mass 
  fraction $m_{\rm strip} = M_0 - M$. Two Milky Way type galaxy models are shown: The 
  LM4 model (used in the SPH simulations by \citet{jachym07}) is rather resistant to 
  stripping because of its radial scale for the gas had been assumed to be as short 
  (4 kpc) as that of the stellar disk. The LM10 model has a longer gas scale (10 kpc) 
  which is closer to observed HI profiles \citep{bigiel12} while having the same total 
  gas mass as LM4. Its lower gas surface density makes the object substantially more 
  vulnerable to ram pressure stripping. To illustrate the effects also on less massive 
  galaxies, we take the example of NGC~4522 which has a rather low rotational velocity 
  near 100 km/s \citep{vollmer00} that makes it more vulnerable. 
     
  The results can be summarized in these terms:
  \begin{itemize}
    \item The gravitational potential of the stellar disk strongly influences the galaxy's response 
          to ram pressure stripping. Due to the potential's strong variation with height above 
          the plane it is the dominant factor which controls the vertical restoring force, 
          which in the long-pulse limit must be overcome. 
          The mass of the dark matter halo contributes to the depth of the gravitational potential 
          and hence to the work that must be done by a gas element to escape. However, since the halo 
          contribution has a scale length larger than the thickness of the disk, it does not much
          affect the vertical restoring force.
    \item The mass and the radial distribution of gas in the disk are important parameters 
          which determine the outcome of a stripping event. Firstly, the conditions for
          stripping (Eqns.\ref{e:shortpulse} and \ref{e:longpulse}) depend on the gas 
          surface density at a certain position in the disk. Secondly, they determine
          the gas mass which is beyond the stripping radius and will be removed.
    \item The bulge determines the gravitational potential predominantly in the region 
          close to the galactic centre. Thus its mass and size affect the stripping outcome 
          only in severely and almost completely stripped galaxies.
    \item The dark matter halo affects mainly the outer parts of the gas disk,
          hence it applies mainly to very weakly stripped objects. 
  \end{itemize}

\section{Predictions For Other Galaxies}\label{s:others}

  The long- and short-pulse limits for a given galaxy to lose a given mass fraction of its 
  gas are described by the maximum ram pressure $p_{\rm max}$ and its time-averaged value 
  $(v\Sigma)_{\rm ICM}$, respectively. Thus they form two straight lines in the diagram of 
  the $(v\Sigma)_{\rm ICM}$ - $p_{\rm max}$ plane. They intersect at some crossing point 
  (see Fig.\ref{f:theplane}) which can be used to characterize the set of limits. As this
  position depends on the properties of the galaxy and the given stripped mass fraction, 
  it is of interest to identify which parts of the diagram are accessible for typical galaxies.

\begin{figure}
  \includegraphics[width=0.33\textwidth, angle=270]{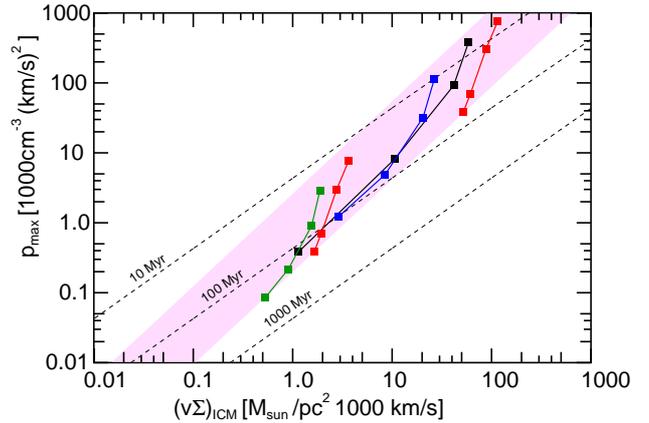}
  \caption{
    Position of the crossing points of short- and long-pulse limits in the
    $(v\Sigma)_{\rm ICM} - p_{\rm max}$ plane for various models:
    Black symbols refer to the LM4 galaxy with squares marking stripped 
    mass fractions of 0.1, 0.5, 0.9, and 0.99 (from lower left to upper right).
    Blue symbols are for NGC~4501 (stellar mass $1.3\,10^{11}$  M$_\odot$), 
    green symbols for NGC~4522 (stellar mass $10^{10}$ M$_\odot$), using 
    Vollmer's respective mass models.
    Red symbols are for a nearly flat gas disk (50 kpc radial scale) in 
    the gravitational potential of a stellar disk of $6.5\, 10^{10}$ (left) 
    and $6.5 \, 10^{11}$ M$_\odot$ (right) without bulge or dark halo.
    The underlying pink area is the region where the crossing points of
    long- and short-pulse limits of usual galaxies are found.
    Dashed diagonal lines indicate the durations of the pressure pulse.}
  \label{f:corridor}
\end{figure}

  When one places a variety of galactic models with realistic parameters and diverse
  values for the stripped mass fraction in the diagram, it 
  becomes apparent that the positions of the crossing points are confined to a narrow region. 
  In Fig.\ref{f:theplane} one notes that the crossing points for the three values 
  of the stripped mass fraction lie on nearly a straight line with a slope somewhat 
  steeper than the lines of equal pulse duration. Since a crossing point describes 
  uniquely the position of its associated curve in this figure, it suffices to display 
  this point in the plane for a given set of galactic parameters and a specified stripped 
  mass fraction, as done in Fig.\ref{f:corridor}. The black squares -- at the same positions 
  as in Fig.\ref{f:theplane} -- occupy a rather narrow band in the plane, which is shared 
  with galaxies of different parameters.

  The position of the crossing point depends strongly on the mass of the galaxy and to a 
  lesser extent on the stripped mass fraction: Red symbols in Fig.\ref{f:corridor} 
  pertain to a galaxy model consisting only of a stellar disk with flat rotation curve and 
  a nearly constant-density gas disk (radial scale 50 kpc). A ten-fold increase of the disk 
  mass requires a hundred times greater maximum ram pressure and $(v\Sigma)_{\rm ICM}$ 
  higher by a factor of $10^{1.5}$ to get the same amount of stripping. To increase 
  the stripped mass fraction from 0.1 to 0.99 one needs a ten times higher $p_{\rm max}$ 
  but only about a factor of 2.5 in $(v\Sigma)_{\rm ICM}$.

  The slope of the overall relation $p_{\rm max} \propto (v\Sigma)_{\rm ICM}^{4/3}$ 
  (pink band in Fig.\ref{f:corridor}) can be understood by using the following 
  approximative relations: The long-pulse limit (Eq.\ref{e:longpulse}) can be written as
  \begin{equation} \label{e:longpulsesimp}
    p_{\rm max} \approx \Sigma_g {v_{\rm circ}^2 \over 2r}  
  \end{equation} 
  as the maximum restoring force can be approximated by the centrifugal force at the outer 
  radius $r$ of the gas disk (cf. Sect.\ref{s:obsass}). The short-pulse limit 
  (Eq.\ref{e:shortpulse}) can be written as
  \begin{equation} \label{e:shortpulsesimp}
     (v\Sigma)_{\rm ICM} = \Sigma_g v_{\rm circ} \sqrt{(v_{\rm esc}/v_{\rm circ})^2 - 1} 
  \end{equation}
  As numerical evaluation of the potentials of complete model galaxies 
  (with $v_{\rm circ}(r) \approx const.$) confirms, we may also assume that 
  $v_{\rm esc}(r) \approx const.$ This yields
  \begin{eqnarray} 
    p_{\rm max}         & \propto & \Sigma_g  r^{-1}\\  
    (v\Sigma)_{\rm ICM} & \propto & \Sigma_g 
  \end{eqnarray} 
  For a weakly stripped galaxy, only the outer parts of the gas disk are affected. Here, 
  the gas surface density profile of a Miyamoto-Nagai disk can be approximated by
  $\Sigma_g(r) \propto r^{-3}$, so that $r \propto \Sigma_g^{-1/3}$, giving  
  the simple relation 
  \begin{equation} 
     p_{\rm max} \propto (v\Sigma)_{\rm ICM}^{4/3}
  \end{equation} 
  which matches the slope of the band depicted in Fig.\ref{f:corridor}. The band's lower 
  boundary constitutes the limit of negligible stripping. It can be shown that no model 
  with physically sensible parameters can have its crossing point in the region below 
  this boundary line. On the other hand, the upper left hand area (pulse duration $<1$ Myr) 
  could only be reached by the crossing points for low mass gas disks embedded in a very 
  massive potential ($> 10^{14}$ M$_\odot$), thus not by typical galaxies.   

  More complete galaxy models whose rotation curve is computed from the
  potential of bulge, disk, and dark halo also are restricted to this
  narrow band, as depicted by the examples for NGCs 4501 and 4522.
  Since the gas disks are modelled by Miyamoto-Nagai disks with a radial scale
  of 10 kpc, the variation to cover stripped mass fractions from 0.1 to 0.99
  is larger in $(v\Sigma)_{\rm ICM}$ than in constant-density disks.
  The LM4 model - due to its shorter gas scale of 4 kpc - shows an even 
  larger spread, because of the greater variation of the gas density in the disk.

\section{Comparison with SPH Simulations}

  How well does the purely kinematical treatment of the gas in terms of parcels 
  fit to the complex gas dynamics involved in ram pressure stripping of a galaxy? 
  This needs a comparison with numerical models including the relevant processes
  in detail. There is no numerical study available that covers a large parameter space
  suitable for comparison with the analytical predictions in Fig.\ref{f:theplane}.
  The closest choice is the SPH simulations by \citet{jachym07} who studied the effects 
  of ram pressure stripping on a Milky Way-like galaxy which radially falls into clusters 
  having a wide range of properties, i.e. the density and the spatial distribution of 
  the ICM. 

  We briefly summarize the essential details of these simulations. The four components 
  of the galaxy -- stellar bulge, stellar disk, gas disk, dark matter halo -- are 
  represented with a total number of 42~000 particles, initially distributed 
  in space according to density profiles described in Sect.\ref{s:galaxy}. The ICM is 
  represented by 120~000 SPH particles distributed following the $\beta$-profile 
  for a model of the Virgo cluster (Sect.\ref{s:virgocluster}). The galaxy 
  starts at rest from the cluster periphery ($R=1$ Mpc) and freely falls on a 
  radial trajectory towards the cluster centre. In the standard run the galaxy reaches 
  the centre after time $T=1.64$ Gyr with a velocity of about 1300 km/s.
  To treat various cluster environments, the standard values for the ICM 
  of $R_{\rm c,ICM} = 13.4$~kpc and $\rho_{\rm 0,ICM} = 6.5\ 10^{-3}\density$ 
  are multiplied with factors of 8, 4, 2, 1, 0.5, and 0.25. The standard run
  yields a maximum ram pressure $p_{\rm max} = 11000\press$
  and $(v\Sigma)_{\rm ICM} = 7000\vsigma$.

\subsection{Cluster model} \label{s:virgocluster}

  The Virgo cluster of galaxies is modelled by the distribution of dark 
  matter (DM) and ICM gas. Their volume densities follow
  $\beta$-profiles \citep{cavaliere76, schindler99, vollmer01}: 
  $\rho = \rho_0 (1 + R^2/ R_{\rm c}^2 )^{-3\beta /2}$ with
  $\beta_{\rm ICM}=0.47$ and $\rho_{\rm 0,ICM}= 0.04\density = 4.0\ 10^{-26}$ g~cm$^{-3}$ 
  and $R_{\rm c,ICM} = 13.4$ kpc. The dark matter, which provides the 
  gravitational potential, has $\rho_{\rm 0,DM}=3.8\ 10^{-4}$ M$_\odot\,$pc$^{-3}$,
  $R_{\rm c,DM} = 320$ kpc, and $\beta_{\rm DM}=1$. 

\subsection{Galaxy model} \label{s:galaxy}

  In what follows we shall focus on a late-type massive galaxy resembling
  the Milky Way (LM-type in \citet{jachym07}): its model is a four-component
  system with Plummer spheres for bulge and dark halo and Miyamoto-Nagai disks
  for stellar and gas disks, with parameters given in Table \ref{t:LMparms}.
  The resultant rotation curve is flat at about $225\speed$.
  We use two versions, LM4 and LM10, which have different radial scales for the gas disk.  

\begin{table}
  \centering
  \caption{The parameters of the components of the LM4 and LM10 galaxy 
    models: mass inside the truncation radius $r_{\rm trunc}$, radial scale $a$ 
    and scale height $b$.}
  \label{t:LMparms} 
  \begin{tabular}{lcccc}
    \hline
                &  mass           & $r_{\rm trunc}$ & $a$ & $b$\\
                &  [M$_\odot$]    & [kpc] & [kpc] & [kpc]\\
    \hline \noalign{\rule{0pt}{0.6ex}}
bulge           & $1.3\,10^{10}$  & 4     & 0.4   & --       \\
stellar disk    & $6.5\,10^{10}$  & 16    & 4     & 0.25     \\
LM4: gas disk   & $6.5\,10^{9}$   & 16    & 4     & 0.25     \\   
LM10: gas disk  & $6.5\,10^{9}$   & 16    & 10    & 0.25     \\   
dark halo       & $3\,10^{11}$    & 20    & 40    & --       \\
    \hline
  \end{tabular}
\end{table}

\subsection{The Stripped Gas}

\begin{figure}
  \includegraphics[width=0.33\textwidth, angle=270]{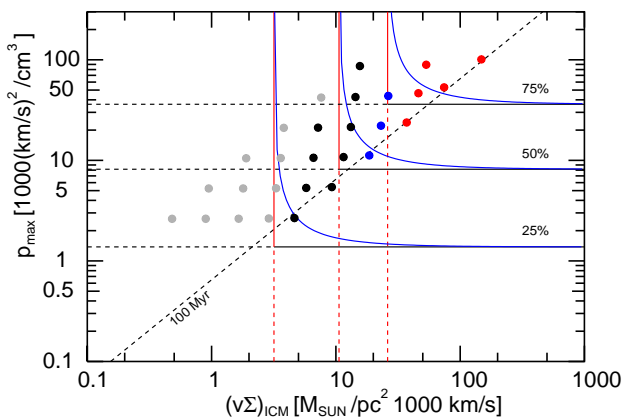}
  \caption{
    Like Fig.\ref{f:theplane}, but compared with the results from SPH models 
    \citep{jachym07} which are shown by large dots whose colour indicates the 
    stripped mass fraction: grey below 0.25, black below 0.5, blue below 0.75, 
    and red above 0.75.}
  \label{f:sphplane}
\end{figure}

  Figure \ref{f:sphplane} shows that the analytical predictions in the 
  $(v\Sigma)_{\rm ICM}$-$p_{\rm max}$ plane agree very well with the results from the
  SPH simulations. One notes that models with the higher values of $(v\Sigma)_{\rm ICM}$ 
  tend to be more strongly stripped than expected from the analytical considerations. This may 
  indicate that in situations of weak and moderate stripping a kinematic description of the 
  event is well sufficient, but that in longer and more severe interactions the hydrodynamic 
  aspects become more important.

\begin{figure}
  \includegraphics[width=0.33\textwidth, angle=270]{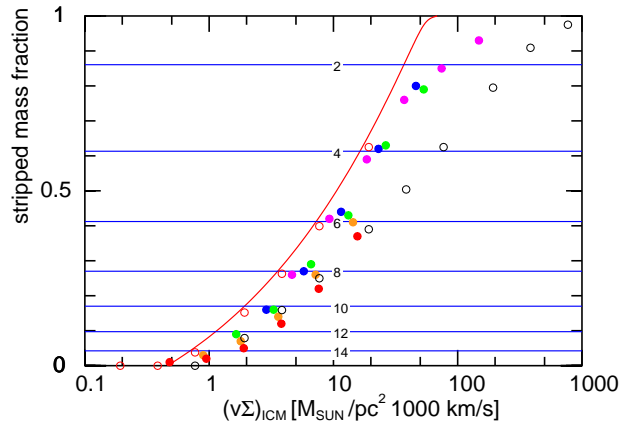}
  \caption{
    The stripped mass fractions as a function of 
    $(v\Sigma)_{\rm ICM}$ predicted analytically from the
    short-pulse limit (red curve) compared to the results from 
    the SPH simulations by \citet{jachym07} and \citet{jachym09}. 
    The colour of the dots indicates the core radius $R_{\rm c,ICM}$ 
    of the ICM distribution: red, orange, green, blue, and magenta refer 
    to 3.4, 6.7, 13.4, 26.8, and 53.6~kpc which correspond to pulse
    FWHM durations of 6, 12, 25, 50 and 100~Myr. Small open circles
    are from test particle models with Gaussian shape force pulses
    of FWHM durations of 2.5~Myr (red) and 250~Myr (black). 
    Horizontal blue lines indicate the outer radius (in kpc) of the remaining gas disk.}
  \label{f:sphplot}
\end{figure}

  A more detailed comparison with the predictions from the short-pulse limit
  (Fig.\ref{f:sphplot}) shows that the SPH results agree very reasonably.
  However, there is a systematic trend: models with the shortest pulses 
  (6 Myr i.e. $R_{\rm c,ICM} = 3.4$~kpc) are not as strongly stripped as predicted 
  by the short-pulse limit, while those with longer pulses agree better. 
  The SPH results form a rather narrow relation, although their durations cover the 
  range from about 6 to 100 Myr, which would result in a larger spread, judging from 
  the test particle models. This concentration is merely due to a systematic change of the 
  pulse shape in the SPH models: Because of computational economy, the ICM particles 
  had been placed only in a sphere of 140 kpc radius about the cluster centre, as to
  cover well the principal phase of the stripping. In a small cluster this would cover 
  indeed a large part of the pressure pulse, which is well approximated by a modified 
  Lorentzian profile ($p(t) \propto (1 + (\Delta t/\tau)^{1.3})$ with the time $\Delta t$
  since maximum pressure and the timescale $\tau$). But at larger clusters the pulse is truncated 
  and its core resembles more a Gaussian. As a consequence, the value of $(v\Sigma)_{\rm ICM}$ 
  is reduced by as much as a factor of 0.6 for the largest cluster, shifting the long-pulse 
  points to the left and creating this narrow band.

\begin{figure}
  \includegraphics[width=0.33\textwidth, angle=270]{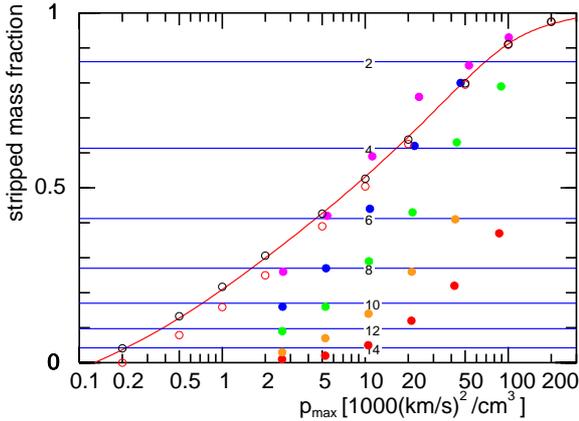}
  \caption{
    Similar to Fig.\ref{f:sphplot}, but as a function of maximum ram pressure 
    (red curve) predicted for the long-pulse limit. The results from the SPH simulations 
    by \citet{jachym07} are depicted with dots whose colour indicates the core radius 
    $R_{\rm c,ICM}$ of the ICM distribution. The small open circles are test particle 
    models with Gaussian shape force pulses of FWHM durations of 2500~Myr (black) and 
    250~Myr (red).}
  \label{f:sphpmaxplot}
\end{figure}

  To show the behaviour close to the long-pulse limit, the stripped fractions are 
  plotted against maximum ram pressure (Fig.\ref{f:sphpmaxplot}). As expected, test 
  particle models with pulse durations longer than about 250~Myr match the analytical 
  relation very well, as was already seen in Fig.\ref{f:sphplane}. The SPH simulations 
  with an ICM core radius $R_{\rm c,ICM} > 20$~kpc, which correspond to pulse widths 
  of more than 50 Myr, are still well represented by the long-pulse limit. 
  For $p_{\rm max} = 10000 ... 50000\press$ the stripped mass fraction 
  exceeds the long-pulse limit. It appears likely that this seeming violation of the 
  long-pulse limit has a similar origin to what is seen in test particle models: due 
  to radial motions gas parcels tend to escape to positions where the gravitational 
  potential is less deep than where the gas had been before the start of the interaction.
  Other reasons for the deviations between SPH models and the analytical prediction
  could be that in the SPH simulations the galaxy evolves, forming instabilities and 
  spiral arms, and the hydrodynamics by which one parcel of gas affects its 
  neighbourhood. 

  In summary, it is remarkable that the purely kinematic considerations alone are 
  capable of predicting stripped mass fractions to within 5\% for face-on 
  stripping events for both short and long interactions, despite the apparently large 
  underlying physical differences between SPH simulations and analytical predictions.

\subsection{The Re-accretable Gas}

  During a stripping event some part of the gas is pushed away from its initial position
  in the galactic plane, but remains gravitationally bound and will eventually fall back
  into the disk. In order to distinguish between the gas that is only slightly perturbed
  and remains close to the galactic plane and gas that is pushed away from the disk,
  we shall define the disk as the cylindric volume within 16~kpc galactocentric radius 
  and within $h=1$~kpc of the plane, similar to \citet{vollmer01}. Gravitationally
  bound gas outside this volume is considered as re-accretable gas. 

  In the short-pulse limit this relation
  \begin{equation} \label{e:shortkick}
    (v\Sigma)_{\rm ICM} = \int p(t) dt \ge \Sigma_g(r) \sqrt{2(\Phi(r,h)-\Phi(r,0))} 
  \end{equation}
  determines the galactocentric distance $r$ outside which gas is kicked out of this volume
  by an event with given $(v\Sigma)_{\rm ICM}$. Integration from this radius outward gives 
  the gas mass fraction $m_{\rm kick}$ which also includes the gas that can also escape. 
  The mass fraction that will eventually be re-accreted is 
  $m_{\rm reacc} = m_{\rm kick} - m_{\rm strip}$, with the stripped part $m_{\rm strip}$ 
  computed from Eq.\ref{e:shortpulse}.

\begin{figure}
  \includegraphics[width=0.33\textwidth, angle=270]{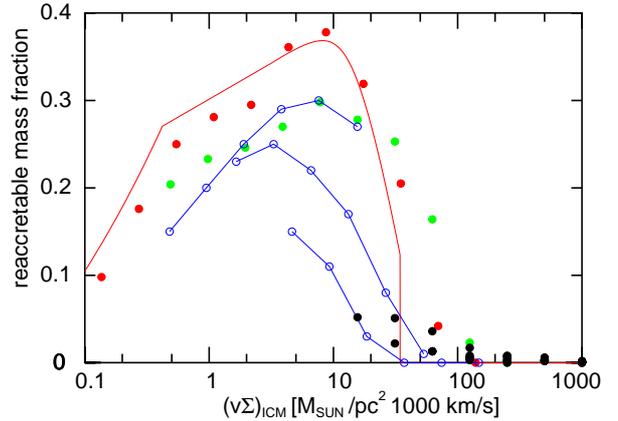}
  \caption{
    The reaccretable mass fraction as a function of $(v\Sigma)_{\rm ICM}$, 
    as predicted from the short-pulse limit (red curve). SPH models are 
    shown as blue circles, with lines connecting models of the same pulse 
    duration: 6,  25, and 100 Myr (from top to bottom). Test particle
    models of 6 Myr duration are shown with Gaussian pulse (red dots) and 
    modified Lorentzian pulse (green dots). Black dots pertain to particle 
    models with pulse durations longer than 200 Myr.}
  \label{f:kickplot}
\end{figure}

  The SPH simulations yield somewhat less reaccretable gas than estimated from the 
  analytical arguments (Fig.\ref{f:kickplot}). Test particle models reveal that 
  the reason is the shape of the pressure pulse: While Gaussian pulse models match 
  the analytical predictions very well, modified Lorentzian pulses give smaller 
  reaccretable mass fractions, as do the SPH models.  

\begin{figure}
  \includegraphics[width=0.33\textwidth, angle=270]{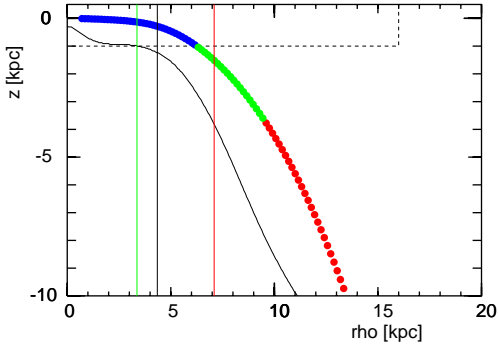}
  \caption{
    The variation (black curve) of the height of the maximum vertical 
    restoring force with distance from the galactic rotation axis, for 
    the LM4 model. The position of the gas parcels from a test particle 
    model with $R_{\rm c,ICM} = 13.4$~kpc, 
    $\rho_{\rm 0,ICM} = 6.5\ 10^{-3}\density$ at the time of maximum 
    ram pressure is shown by coloured dots: blue dots are parcels that remain
    in the disk (the volume marked by a dashed line) green dots are those 
    kicked out of that volume but remaining gravitationally bound to the galaxy, 
    and red dots mark those already escaping. Vertical lines indicate for this 
    standard run the radii outside which gas elements are expected to be 
    stripped in the limits of short (red) and long (black) ram pressure pulses. 
    The green line indicates the radius beyond which gas elements are
    kicked out of the disk volume, estimated from the short-pulse limit.
    After the pulse, the green line will indeed form the border between blue and 
    green dots, and the red line will separate green and red dots.}
  \label{f:frestkicked}
\end{figure}

  Figure \ref{f:frestkicked} shows the situation at maximum pressure for the test 
  particle model equivalent to the standard SPH run. The reaccretable mass extends 
  up to 4~kpc away from the galactic plane, where it joins the material already escaping. 
  After the end of the pressure pulse, this border at 10 kpc galactocentric radius 
  will have further receded to 7~kpc, and the inner border of the kicked material 
  will be at 3.3~kpc radius, as predicted from the short-pulse limit. This leaves 
  an appreciable mass fraction that will eventually fall back into the disk. 

  For the outcome in the long-pulse limit one notes from Fig.\ref{f:frestkicked} 
  that for the LM4 model galaxy the height of peak restoring force is  
  close to 1~kpc for the intermediate part of the disk. This means that once 
  a gas parcel is pushed up to this height, the continued action by the long 
  pulse will drive it into escape. Hence, with our choice for the height of 
  1~kpc for the disk proper, there will remain no or little reaccretable gas.
  This is confirmed in the test particle models and SPH models with pulse durations 
  (50 and) 100~Myr and large values of $(v\Sigma)_{\rm ICM}$. SPH models with low 
  values of $(v\Sigma)_{\rm ICM}$ show appreciable reaccretable mass fractions. 

\section{Orbits in Clusters}\label{s:orbits}

  As discussed in Section \ref{s:combine}, the stripping outcome for
  a given galaxy is determined by the position in the
  $(v\Sigma)_{\rm ICM}$--$p_{\rm max}$ plane. In the idealized situation of 
  a spherically symmetric ICM density distribution about a single cluster centre
  the peak ram pressure occurs at the galaxy's closest approach to the cluster 
  centre. The shape of the orbit influences the duration of the pressure pulse.
  Thus, the position in the $(v\Sigma)_{\rm ICM}$--$p_{\rm max}$ plane is also
  associated with the parameters of the galaxy's orbit. This allows us to determine 
  which kinds of orbits in a given cluster would be suitable to cause stripping
  in a given galaxy.  

\subsection{The Most Likely Trajectories}

  When a galaxy follows its path through the cluster, the orbit is determined by the 
  cluster's gravitational potential. The maximum ram pressure is reached at the 
  orbit's pericentre and its value is determined from the velocity -- hence essentially 
  by the cluster distribution of dark matter -- and the local ICM density. 
  The rosette-type orbit can be characterized by the ratio of its apo- to pericentric 
  distances $r_{\rm a}/r_{\rm p}$ and its closest approach to the centre $r_{\rm p}$.
  High-resolution cosmological N-body simulations suggest that the orbits of infalling 
  satellite haloes \citep{wetzel11} -- cf. the dependence on host halo mass in their 
  Fig.4 -- have mostly orbital circularities around 0.5, i.e. an eccentricity of 0.87, 
  which gives an axial ratio $r_{\rm a}/r_{\rm p} \approx 14$. 
  From their Fig.A1 we take that the most likely pericentric distance is 
  $r_{\rm p} \approx 0.2 r_{\rm vir}$ of the virial radius.

\subsection{Virgo as an Example}

\begin{figure}
  \includegraphics[width=0.33\textwidth, angle=270]{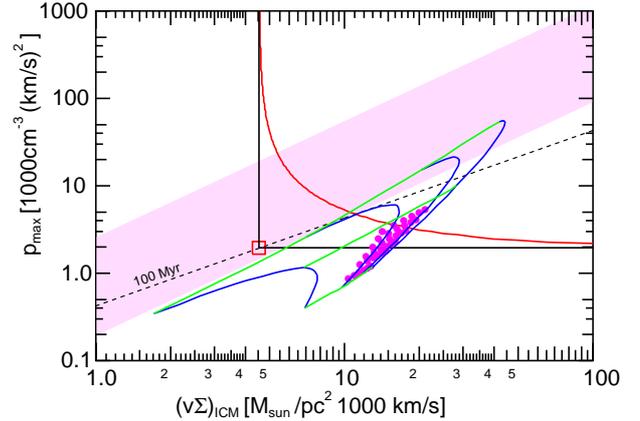}
  \caption{
    Loci in the plane of maximum $p_{\rm max}$ and time-integrated ram pressure 
    $(v\Sigma)_{\rm ICM}$ predicted for orbits through the Virgo ICM as modelled 
    by \citet{schindler99}: Green lines mark minimum radii of 0.01, 0.1, and 0.5 Mpc 
    (from left to right), blue curves depict $r_{\rm a}/r_{\rm p} = 3, 10, 30, 100$ 
    (from bottom to top). Magenta dots mark the more likely orbits in the range of  
    $7 < r_{\rm a}/r_{\rm p} < 28$ and $110 < r_{\rm p} < 440$ kpc. Horizontal and 
    vertical black lines are the long- and short-pulse limits for the LM10 galaxy 
    model and a stripped mass fraction of 0.5, with the red interpolating curve. 
    As in Fig.\ref{f:corridor}, the underlying pink area marks the region where 
    the crossing points of long- and short-pulse limits for usual galaxies are found.
    The diagonal dashed line indicates a pulse duration of 100 Myr.}
  \label{f:virgoplane}
\end{figure}

  Using simple calculations we explore orbits with parameters $r_{\rm a}/r_{\rm p}$ 
  and $r_{\rm p}$ from ranges (specified in Fig.\ref{f:virgoplane}) which include 
  the most likely values. From each pair of parameters the initial conditions for the 
  orbit are determined, and the flight of a galaxy through the cluster can numerically 
  be computed, which gives the maximum ram pressure $p_{\rm max}$ and its time integral 
  $(v\Sigma)_{\rm ICM}$. The computation is stopped after its pericentric passage when 
  the galaxy reaches twice the pericentric distance. 

  Let us consider a galaxy like the Milky Way in the Virgo cluster. With the parameters 
  given in Sections \ref{s:virgocluster} and \ref{s:galaxy}, one obtains the results presented 
  in Fig.\ref{f:virgoplane}. The area in the $(v\Sigma)_{\rm ICM}$--$p_{\rm max}$ plane 
  for models with these orbits is rather small: For the most likely orbits (magenta dots)
  they cover a range in $(v\Sigma)_{\rm ICM} = 10$ to $20$ and in $p_{\rm max} = 1$ to $5$ 
  in the units of the figure.  

  The reason for such a concentration of the models is that the duration of the pressure 
  pulse is limited to a rather narrow range: Consider a galaxy on its highly eccentric orbit. 
  The path length on the orbit during which the galaxy is subjected to high ram pressure 
  would be of the order of the pericentric distance. Since its speed will be close to the 
  escape velocity $v_{\rm esc}$ from the cluster centre, the duration of the pressure pulse 
  can be estimated as
  \begin{equation}
     \Delta t \approx r_p/v_{\rm esc} = 0.2 \times r_{\rm vir}/v_{\rm esc}
  \end{equation}
  with the cluster's virial radius $r_{\rm vir}$. The escape speed can be estimated 
  from the cluster mass
  \begin{equation}
     v_{\rm esc} = \sqrt{{G M \over  r_{\rm vir}}} 
                 = \sqrt{{4\pi\over 3} G\rho_{\rm DM}} \,\, r_{\rm vir}
  \end{equation}
  with the mean dark matter density $\rho_{\rm DM}$. This results in a pulse duration
  \begin{equation}
     \Delta t \approx 0.2 \times \sqrt{{3\over 4\pi G \rho_{\rm DM}}}   
  \end{equation}
  which is a multiple of the free-fall time scale of the cluster. 

\subsection{Other Clusters}

\begin{figure}
  \includegraphics[width=0.33\textwidth, angle=270]{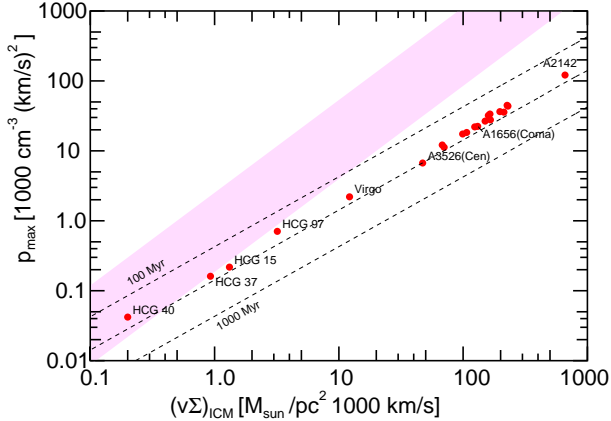}
  \caption{
    Similar to Fig.\ref{f:virgoplane} but with a dot for the most likely
    orbit ($r_p = 0.2 r_{\rm vir}$, $r_a/r_p = 14$) representing each cluster. 
    Dashed lines indicate the duration of the pressure pulse.}
  \label{f:clusters}
\end{figure}

  The conditions for ram pressure stripping in a cluster can thus be characterized by 
  the most likely orbit, which may be computed from a mass model for the cluster.
  For a number of galaxy clusters, the X-ray observations have been modelled by 
  \citet{mohr99} and \citet{ettori02}, yielding parameters for the distributions of ICM 
  and dark matter. These models show no trend of the mean dark matter density with 
  total mass. Numerical simulations of galactic orbits with $r_p/r_{\rm vir} = 0.2$ and 
  $r_a/r_p = 14$ give FWHM pulse durations in the range of 150 to 400 Myr with a strong 
  concentration near 200..300 Myr. As a consequence, clusters form a very well-defined 
  relation $p_{\rm max} = 0.2 (v\Sigma)_{\rm ICM}$ in the plane of maximum and time-integrated 
  pressure (Fig.\ref{f:clusters}).
  The much less massive Hickson compact groups \citep{rasmussen08} have somewhat higher 
  densities and require slightly lower pulse durations, but follow the same relation as 
  massive clusters. 

\subsection{Implications for Galaxies in Clusters}

  Figure \ref{f:corridor} shows that the band to which the crossing points of the short- 
  and long-pulse limits for realistic galaxies are restricted is characterized by pulse 
  durations of less than about 300 Myr. Thus, the conditions for the most likely 
  orbits in clusters (Fig.\ref{f:clusters}) are found at higher values of 
  $(v\Sigma)_{\rm ICM}$ (for a given maximum ram pressure) than the crossing point which 
  represents the minimum value for $(v\Sigma)_{\rm ICM}$ for the long-pulse limit. 
  Therefore, the long-pulse limit is the more appropriate approximation to explain stripping 
  events in the clusters. However, especially in low mass clusters and groups, this limit 
  may underestimate the required maximum pressure (cf. Figs.\ref{f:theplane} and 
  \ref{f:virgoplane}).

  The consequences for galaxies can be estimated from Fig.\ref{f:corridor}: 
  An object like NGC~4501 on the most likely orbit in Virgo 
  (with $p_{\rm max} \approx 2000\press$) would lose only 10\% of its gas, but in the 
  harsher environment of Coma ($p_{\rm max} \approx 20000\press$) it would lose about 90\%. 
  A smaller galaxy such as NGC~4522 could lose 90\% in Virgo and would be stripped of all 
  its gas in Coma. A galaxy as robust as the LM4 model would attain stripped mass fractions 
  of 0.2, 0.6, and 0.9 in Virgo, Coma, and A2142 ($p_{\rm max} = 100000\press$, respectively.  
  
  Since pulse durations are concentrated in the range of 200-300 Myr and since 
  galaxies like the Milky Way have periods for vertical oscillations as long as 100 
  to 200 Myr in their outer regions (as seen in Fig.\ref{f:vertperiod}), the long-pulse 
  limit may describe weakly stripped galaxies in these cluster less accurately. 
  In situations when the pulse duration is as short as to approach the period for 
  vertical oscillations at the stripping radius, the criterion of \citet{gunngott72} 
  would underestimate the ram pressure to cause a given deficiency. This is about a 
  factor of 2 or 3, as shown in the example from Fig.\ref{f:virgoplane}.

\subsection{Applicability of the Short-Pulse Limit}

  As we have shown in the previous section, the long-pulse limit provides a good
  approximation for ram pressure stripping of typical galaxies in most clusters.
  There may be some situations when galaxies experience effects of short pressure 
  pulses. 

  When a galaxy moves with a speed of $1000\speed$ relative to the ambient
  medium, the encounter with density enhancements smaller than 10 kpc would result
  in a pulse width of less than 10 Myr. The distribution of ICM in many clusters
  is far from homogeneous. X-ray observations revealed in the Coma cluster ICM 
  linear structures with a width of 10 kpc \citep{sanders13}. \citet{simionescu10}
  identify in Virgo a surface brightness edge as a cold front, i.e. a contact discontinuity 
  associated with gas sloshing in the ICM. The HI tail left by NGC~4388 \citep{oosterloo05} 
  has a width between 15 and 40 kpc. Furthermore, there could be hydrodynamical interactions 
  of galaxies: The recent observations of a ram pressure stripped galaxy UGC 6697 
  and its small companion CGCG 97087N in the Abell 1367 cluster suggest that the small 
  galaxy may have crossed the disk of the main galaxy \citep{consolandi17}.

\section{Comparison with Observations: Virgo Cluster}

  Since our simple, purely kinematical approach is able to reproduce the SPH simulations 
  of ram pressure stripping rather well, we may now attempt to use it to interpret
  observational data from the Virgo cluster. The aim is to compare the ram pressure that 
  a galaxy might experience at its current position -- as estimated from the distribution of
  the ICM which is known from X-ray observations -- to the maximum ram pressure 
  that a galaxy must have undergone, which is obtained from the current state of its gas 
  disk -- its HI radius or its HI deficiency. These two pressure estimates should agree in 
  galaxies that undergo active stripping. Galaxies that had experienced stripping long ago 
  would now be in a position where the ICM density is too low to account for their stripping 
  state. This allows us to classify the galaxies, according to their state as 
  {\it active} and {\it past strippers}. 

  While a complete modelling of a stripping event requires the knowledge of numerous parameters
  that are necessary to describe in detail the galaxy, the cluster, and the galaxy's trajectory 
  through the cluster, it is possible to cut down these requirements substantially.
  
  As shown in Sect.\ref{s:orbits} application of the long-pulse limit appears to be the 
  more appropriate approximation to describe stripping events in galaxies falling into 
  clusters. The criterion of \citet{gunngott72} involves only the maximum ram pressure
  the galaxy has experienced so far, but does not require any information about the detailed 
  time history of the stripping event. As shown by comparison with SPH simulations in a wide 
  range of parameters, application of the analytical criterion for stripping gives reliable 
  estimates for the amount of stripped material. Simulations of various kinds 
  \citep{vollmer01, roediger06, jachym09} have shown that ram pressure stripping of a tilted 
  galaxy is as effective as in the face-on case, as long as the angle between the flight 
  direction and the normal to the disk is less than about $60^o$. Thus an interpretation 
  with the face-on formulae is a reasonable approximation. 
  
  \citet{vollmer01} pointed out that the results of their simulations of face-on stripped 
  galaxies can well be reproduced by applying the criterion of \citet{gunngott72} with 
  the maximum restoring acceleration given by the centrifugal acceleration at the outer 
  rim of its gas disk. In fact, as will be shown in Fig.\ref{f:acenamax} below, the
  maximum restoring force and the centrifugal acceleration are closely related. 
  As the gravitational potential of spiral galaxies can well be represented by their 
  rotational velocity by assuming a flat rotation curve, a detailed modelling of the 
  potential is not necessary.

  The HI gas in spiral galaxies is commonly distributed with a rather flat radial profile, 
  decreasing outwards \citep{bigiel12}. Hence, reasonable assumptions, such as an exponential 
  disk with a certain radial scale, may be viable. 

  As only global properties, such as optical radius $r_{opt}$, rotational velocity $v_{\rm rot}$, 
  observed HI mass $M_{\rm HI}$, and HI deficiency (in the usual definition as 
  $def = \log_{\rm 10}(M_0/M_{\rm HI})$ where $M_0$ is the galaxy's HI mass before stripping)
  are involved, it is possible to use observational data available for a large number
  of galaxies. In what follows, this is done for the Virgo cluster with data taken from 
  the GOLDMine database (\citet{gavazzi03}, hereafter referred to as GoldMine), which
  provides homogeneous data for a large number of galaxies. Selecting non-elliptical 
  objects with rotational velocities larger than 30 km/s yields a sample of 
  232 objects, of which 197 are HI deficient. Of these are 137 spirals with 
  optical radii between 2 and 20~kpc and 84 dwarf galaxies with radii below 
  about 5~kpc (predominantly of types Im, Sc, Sm, and some BCDs). 
  For 39 larger objects additional information, such as the HI radius, are available from 
  VIVA \citep{chung09}. The results from detailed modelling for the objects in various phases 
  of stripping by \citet{vollmer09} can serve as a control sample with known stripping properties.

\subsection{Assumptions}\label{s:obsass} 

\begin{figure}
  \includegraphics[width=0.33\textwidth, angle=270]{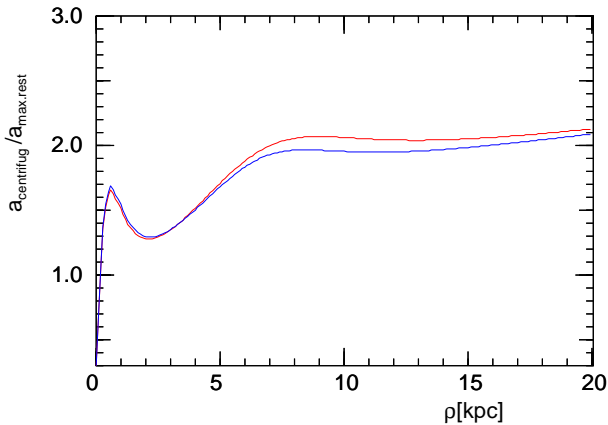}
  \caption{
    The ratio of centrifugal acceleration and maximum restoring vertical acceleration
    as a function of distance from the galactic centre in the LM4 model 
    (red, data from Table \ref{t:LMparms}) and the model for NGC~4522 by 
    \citet{vollmer00} (blue).}
  \label{f:acenamax}
\end{figure}

  In both the foregoing analytical considerations and the SPH models the ISM is treated 
  as a single phase gas, which resembles neutral hydrogen. The comparison with observation
  also concentrates on the easily observable atomic hydrogen, which provides most of the 
  information about stripped galaxies in the form of the HI deficiency. As the ISM is 
  a complex mixture of neutral, molecular and ionized phases at various temperatures,
  the removal of neutral hydrogen from the disk is also influenced by the presence of 
  the other components. Numerical simulations of the response of a multi-phase ISM to a 
  stripping event (e.g. \citet{tonnesen09}) show that diffuse HI can easily leave the disk, 
  and HI is also removed by ablation from envelopes of denser clouds which have molecular cores. 
  However, these models do not yet lend themselves to derive generally valid prescriptions 
  of how the presence of the molecular fraction affects the HI mass loss. 
  Thus we may use the empirical approach by \citet{vollmer01} who added to the restoring 
  force an enhancement factor which takes into account that in the inner parts of a spiral 
  galaxy the presence of molecular cores make it more difficult to remove neutral hydrogen 
  from gas clouds: The restoring force is multiplied by 
  \begin{equation}
    \label{e:retain}
    f  =  (1 + a \exp(-r_{\rm strip}/R_0))       
  \end{equation}
  where $a = 15$ is the enhancement factor and $R_0 = 2$~kpc a radial scale associated with 
  the molecular gas fraction \citep{vollmer01}. Thus the stripping radius $r_{\rm strip}$ -- 
  defined as the outer edge of the HI gas that remains gravitationally bound to the galaxy -- 
  is determined by the ram pressure $p$ being equal to the maximum restoring force: 
  For a HI gas parcel with surface density $\Sigma_{\rm HI}(r)$ at distance $r$ from the 
  centre of the disk this occurs where this condition holds:
  \begin{equation}
      p =  f \times \Sigma_{\rm HI}(r_{\rm strip}) 
           \bigg|{\partial \Phi(r,z)\over \partial z}\bigg|_{\rm max} 
  \end{equation}
  If we were to use for galaxies of all sizes the fixed value of the radial scale $R_0 = 2$~kpc, 
  the ram pressure would tend to be overestimated in small galaxies. \citet{chung17} find that 
  the diameters of the CO emission are well correlated with the optical diameters, independent 
  of the environment of the galaxies. Their Fig.7 suggests that the majority of their objects 
  has $D_{\rm CO} = 0.75\, D_{\rm opt}$. Therefore we scale $R_0$ with the galaxy's optical 
  radius $R_{\rm opt}$: 
  \begin{equation}
      \label{e:r0scaling}
      R_0 = 2 \,{\rm kpc} \times {R_{\rm opt} \over 15 \,\,{\rm kpc}}
  \end{equation}
  The expression for the stripping radius can be simplified, as the maximum restoring force 
  at some distance but perpendicular to the plane of rotation is closely related to the 
  centrifugal acceleration at this distance. For the gravitational field around a point mass, 
  the ratio $g$ of centrifugal force and maximum restoring force is $\sqrt{2}\,\, 1.5^{3/2} = 2.598$. 
  For Plummer spheres and Miyamoto-Nagai disks the ratio is zero at the centre and rises to converge 
  to this value for distances larger than the radial scale parameter. In the total gravitational 
  potential of various models for disk galaxies we find that the ratio in the outer parts always 
  remains nearly constant, as shown in Fig.\ref{f:acenamax} for two examples. In the following 
  we adopt the value of $g = 2.0$ for this geometrical factor.
  Lastly, with a flat rotation curve at speed $v_{\rm rot}$, the centrifugal acceleration at the 
  outer radius $r_{\rm strip}$ of the HI disk is:
  \begin{equation}
      a_{\rm max} = {v_{\rm rot}^2 \over r_{\rm strip}}
  \end{equation}
  Thus, a complete description of the gravitational potential of the galaxy is not needed, as it 
  suffices to use the observed value for the rotational speed.
  
  The HI surface density at the stripping radius is obtained from a reasonable guess 
  for the initial spatial distribution of HI in the galaxy: Let the surface 
  density follow a Miyamoto-Nagai profile as a function of the distance $r$ from the 
  galactic centre  
  \begin{equation}
     \Sigma_{\rm HI}(r) = \Sigma_0 \, (1 +(r/R)^2)^{-3/2}  
  \end{equation}
  with its value $\Sigma_0$ at the centre and the radial scale $R$. For an initial
  outer radius $r_{\rm max}$ one has
  \begin{equation}
     \Sigma_0 = {M_0 \over 2\pi R^2 (1-(1 + (r_{\rm max}/R)^2)^{-1/2})}  
  \end{equation}
  from the initial mass $M_0$ in the HI disk, which is computed from the galaxy's present 
  deficiency and HI mass by $M_0 = M_{\rm HI}\, 10^{def}$. Other gas distributions, such as 
  an exponential profile, lead to similar expressions. As shown in Fig.\ref{f:defRRatio}, 
  the galaxies in the Virgo cluster can be well described by having the HI disk of a radial 
  scale equal to the optical radius $R = r_{\rm opt}$ and an initial outer radius 
  $r_{\rm max} = 1.5 R$. 

  If the stripping radius is known from HI maps (which pertains to the subsample
  of galaxies in VIVA), a direct estimate for the ram pressure is then obtained from
  \begin{equation}
    \label{e:pmax}
     p  =  {\Sigma_0 \over (1 +(r_{\rm strip}/R)^2)^{3/2}} 
           {v_{\rm rot}^2 \over g \,\,r_{\rm strip}}
           \, (1 + a\, e^{-r_{\rm strip}/R_0})       
  \end{equation}
  Since this is the maximum value that the galaxy must have experienced so far on its
  flight through the cluster, we shall name this estimate the 
  {\it maximum ram pressure} $p_{\rm cfg}$, as it is derived from the centrifugal 
  acceleration. 

  In most objects of our full sample the outer radius of the HI disk cannot be measured.
  But with the same assumptions about the spatial HI distribution one has
  \begin{equation}
     \label{e:massratio}
     10^{-def} = {m(r_{\rm strip}) \over m(r_{\rm max})} = 
                        {1 - 1/\sqrt{1 + (r_{\rm strip}/R)^2}\over 
                         1 - 1/\sqrt{1 + (r_{\rm max}/R)^2}}  
  \end{equation}
  from which the stripping radius for a given deficiency is obtained.   
  Using this value in Eq.\ref{e:pmax} gives the estimated pressure necessary to produce the 
  deficiency. This estimate we call the {\it required ram pressure} $p_{\rm def}$.

  A comparison with the SPH calculations of \citep{jachym07} shows that for the models
  with large ICM core radius (53.6 kpc) application of Eq.\ref{e:pmax} (but with $a=0$) 
  recovers the stripping radius and the maximum ram pressure within 20\%.

  To facilitate the estimation of ram pressure effects on arbitrary galaxies
  we have written a few interactive JavaScript tools, 
  at http://www.astrophysik.uni-kiel.de/$\sim$koeppen/JS/RPShome.html

  The ICM density at the galaxy's projected position from the cluster centre (from 
  the model of \citet{schindler99} with parameters given in \citet{vollmer01}) and 
  the local escape velocity (as an estimate of its flight velocity) yield an independent 
  estimate for the actual ram pressure, which we shall name {\it local ram pressure} $p_{\rm loc}$. 
  As this value is computed for the projected distance from the cluster centre, which 
  is the minimum distance where the galaxy might be, the true local ram pressure would 
  be less for galaxies that lie in the fore- or background in a region of lower density 
  further away from the cluster centre than the projected distance. 
  Since both assumptions are upper limits to the galaxy's speed and the ICM density
  the values constitute strict upper limits for the local ram pressure.

\subsection{Uncertainties of Ram Pressure Estimates} 
 \label{s:uncertain}

  The estimates $p_{\rm cfg}$, $p_{\rm def}$, and $p_{\rm loc}$ for the ram pressure acting 
  at a galaxy are subject to several uncertainties from measured quantities and from assumptions
  for the modelling.
  
  For the objects common to VIVA and GoldMine, optical diameters agree within a standard 
  deviation of 0.08 dex, HI masses by 0.1 dex, and widths of the HI line by about 
  $15\speed$. In this paper we compute HI deficiencies with the formula from \citet{gavazzi13}; 
  the values derived by VIVA agree within 0.22 (standard deviation). 

  VIVA's isophotal HI diameters are within 0.1 dex of the values by \citet{cayatte90}. 
  This diameter is affected by the presence of extraplanar gas. NGC~4522 is a particularly 
  striking example: The radius at which gas starts to move out of the galactic plane is 3 kpc 
  \citep{kenney04}. With this value, the maximum ram pressure of $1850\press$ is much larger 
  than the local ram pressure of $220\press$. With our modelling of the gas disk, a stripping 
  radius of 4.8 kpc is deduced from the deficiency, yielding a required ram pressure 
  of $p_{\rm def} = 520\press$.

  Rotational velocities are derived from the widths of the HI line, thus they represent 
  the maximum value at the rim of the HI disk. Since in numerous galaxies the rotational 
  velocity still increases with radius -- for example NGC~4569 \citep{guhathakurta88, 
  cortes15} -- the derived maximum ram pressures may tend to be overestimated in galaxies
  whose HI radius is much smaller than its optical radius.

  For the estimated ram pressures one needs to know the HI surface density at the 
  outer edge of the HI disk. Comparison of the values derived from our modelling of the
  HI disk and the azimuthally averaged profiles from VIVA suggests that the uncertainties
  may amount to a factor of 2.

  The interpretation of the data by applying the long-pulse limit might lead to systematic 
  underestimation of the ram pressure by a factor 2 to 3 (cf. Fig.\ref{f:virgoplane}), if 
  the actual pulse length is comparable to the period for vertical oscillations. But as this
  period is longer in the outer parts of the disk, this underestimation would be restricted
  to low deficient objects only. Likewise, the application of the face-on case 
  to objects which undergo tilted ram pressure stripping would lead to systematic 
  underestimation of ram pressure by factors of 2 to 3 (cf. \citet{jachym07}). 
   
  In the analysis the ICM is modelled by a smooth and spherically symmetric density 
  distribution centered on M87, while the observed X-ray emission \citep{boehringer94}
  shows substantial structure, including the presence of a second centre about M49. 
  From their residual map we estimate variations of up to a factor of 20 in projected 
  density. \citet{shibata01} find temperature variations of a factor of 4, on scales 
  larger than about 300 kpc. The density profile is "very smooth" and well-fitted by 
  a single $\beta$ profile except in the southern region. 
  
  Our use of the escape speed considers the limiting case of the infall of galaxies into 
  the cluster occurring on a radial trajectory. But this is not a severe limitation: 
  For the most likely orbit (\citet{wetzel11}, as discussed in Sect.\ref{s:orbits}) with 
  $r_{\rm a}/r_{\rm p} \approx 14$, $r_{\rm p} \approx 0.2 r_{\rm vir}$, and 
  $r_{\rm vir} \approx 1$~Mpc for Virgo one gets $r_{\rm a} \approx 3$~Mpc. The maximum speed 
  of this orbit with pericentric distance of 200 kpc is $1700\speed$, only slightly less 
  than the escape speed of $1770\speed$ from the centre of the $\beta$-model to a distance 
  of 3~Mpc. This would reduce the local ram pressure by about 10 percent.   

  An important problem is the true distance to individual galaxies, in particular since
  the Virgo cluster is known to be of triaxial shape \citep{mei07}. While for almost all 
  objects we use the distances given by GoldMine (17 Mpc for the main body, 23 and 32 Mpc 
  for the infalling clouds), individual distances have been determined for 11 galaxies by 
  \citet{cortes08} based on the stellar kinematics in the inner 2 kpc. 
  As these values may differ considerably, this affects the linear distance to M87 and 
  hence the deduced local ram pressure. A few typical examples are indicated in Fig.\ref{f:PlocPcfg}.
  NGC~4424 is a particularly striking example: GoldMine places it in the 23~Mpc cloud. 
  If this were true and if the 23 Mpc cloud's ICM density were also taken from the model 
  for the main cluster, this galaxy would be classified as a past stripper. But the distance 
  of $15\pm1.9$ Mpc from \citet{cortes08} makes it a member of the main cluster, and it 
  counts as an active stripper, where it should belong because of its gas tail 
  (VIVA, \citet{sorgho17}). Also, four arrows at the bottom of Fig.\ref{f:PlocPcfg} show
  by how much the estimated local ram pressure would decrease, if the distance of galaxies at
  various projected positions would depart by 1 Mpc from the assumed distance of 17 Mpc.
  (from right to left: NGCs 4388, 4405, 4532, and 4808).

  Modelling the enhanced retainment of HI by the molecular phase in the inner parts
  of a galaxy by the enhancement factor from Eq.\ref{e:retain} adds another
  uncertainty, since this factor becomes quite important in galaxies with small HI radii,
  either dwarf galaxies or those with strongly truncated HI disks. The effect of this 
  modelling will be shown in the next Section (Fig.\ref{f:PlocPcfg}).

  It is rather difficult to assess the total error from these various contributions. 
  Taken all together, we estimate that the accuracy of the pressure values can certainly 
  be not better than a factor of 2. 

\subsection{Comparison of Ram Pressure Estimates} 

  In this analysis we compare the {\it local ram pressure} $p_{\rm loc}$ provided by the
  ICM with the pressures that the galaxy must have experienced so far: for objects 
  with measured stripping radius the centrifugal acceleration is used to deduce the 
  {\it maximum ram pressure} $p_{\rm cfg}$. For all galaxies we also use the measured 
  HI deficiency to deduce the stripping radius and thus derive the {\it required ram pressure} 
  $p_{\rm def}$ which accounts for the deficiency. This comparison allows us to distinguish 
  two regimes: 
  \begin{itemize}
    \item Galaxies whose maximum (or required) ram pressure exceeds the local value are 
          objects that had undergone stripping in the past and are now in a region of 
          low ICM density which would no longer cause any further stripping.
    \item If the two pressure estimates are comparable, the galaxy is currently undergoing 
          active stripping. Because the local ram pressure is computed with the projected 
          distance from the cluster centre, objects whose local ram pressure is larger than 
          the maximum (or required) ram pressure are likely to be actively stripping, but 
          in the fore- or background.  
  \end{itemize}

  Let us first consider a subsample of 39 objects from the VIVA catalogue. As these objects 
  have known HI (isophotal) diameters, we can compare the local ram pressure with the 
  {\it maximum ram pressure} $p_{\rm cfg}$ (computed from Eq.\ref{e:pmax}) inferred directly
  from the centrifugal force at the outer edge of the gas disk. 

\begin{figure}
  \includegraphics[width=0.33\textwidth, angle=270]{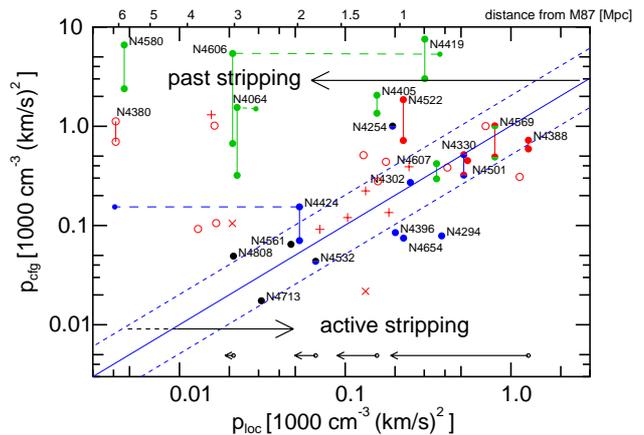}
  \caption{
    The maximum ram pressure $p_{\rm cfg}$ estimated from the current centrifugal force at the outer 
    rim of the HI disks compared to the local ram pressure $p_{\rm loc}$ obtained from the ICM density
    at the projected distance from M87 (indicated by the scale at the top) and assuming that the galaxy
    moves through the ICM with the local escape velocity. The sample comprises only objects with HI isophotal 
    diameter measurements (from VIVA). Red dots mark the sequence of stripped objects \citep{vollmer09}, 
    blue dots are galaxies from \citet{chung07}(with one-sided HI tails), green dots are galaxies with
    clearly truncated HI disks, as found by inspection of HI maps in VIVA, and black dots are 
    non-deficient galaxies. Open circles are HI deficient galaxies ($def > 0.3$), plus signs mark 
    low-deficiency objects ($def < 0.3$), and X-signs are non-deficient objects  ($def < 0$), 
    based on the formula of \citet{gavazzi13} for deficiency. In several objects a lower second
    dot indicates the position if the enhancement factor for the molecular phase is not applied.
    For NGCs 4424, 4606 and 4064 smaller dots indicate the loci if the distances were
    taken from GoldMine instead of the individual distances from \citet{cortes08}. Four arrows 
    at the bottom indicate the decrease of $p_{\rm loc}$ if the distance differed by 1~Mpc from 17~Mpc.
    The full blue line indicates equality of the pressures, the dashed lines a factor of 2 in 
    either sense.}
  \label{f:PlocPcfg}
\end{figure}

  Figure \ref{f:PlocPcfg} shows that near the line of equality one finds galaxies with one-sided 
  tails and extensions (blue dots), which evidently experience ongoing stripping, as well as 
  the galaxies of the sequence of \citet{vollmer09} (red dots), which are in stages near maximum 
  ram pressure. Negative deficiency, gas-rich galaxies (black dots) are also found close to the 
  line of equality but at low pressures. Evidently the conditions at their outer rims are in 
  balance with the local conditions at their large projected distance from M87.

  Most galaxies with well-truncated HI disks (green dots), which are a sign for stripping 
  in the past, have indeed maximum pressures larger than the local value. Their projected 
  distances to M87 are above 1~Mpc, thus they appear to be already on their outward journey 
  through the cluster. NGCs~4064, 4293, and 4580 are identified by \citet{yoon17} as members 
  of the 'backsplash' population \citep{gill05}. Galaxies that had been stripped on their 
  first passage through the cluster centre and had continued on unperturbed highly eccentric 
  trajectories would now be in the outskirts of the cluster. They could be found either at 
  large projected radii with low line-of-sight speeds (such as NGCs~4064, 4293, 4580, and 
  4606 with 394, 386, 275, $333\speed$) or at low projected distances moving nearly parallel 
  along the line-of-sight with higher speeds (such as NGCs~4419 and 4607 with 1580 and 
  $964\speed$). Hence all these objects could be members of this same group.

  The application of the pressure enhancement factor (in Eq.\ref{e:pmax}) due the molecular 
  gas phase can be quite important in some galaxies. In Fig.\ref{f:PlocPcfg} the lower dot of 
  several galaxies indicates the value of $p_{\rm cfg}$ when the factor is not applied. 
  The difference is large in galaxies with HI radii less than about 3~kpc. In most galaxies 
  that currently undergo stripping only those parts are affected which lie well outside the 
  inner region with high molecular content, and hence this factor does not matter. 

\subsubsection{The whole Sample}

  As most of the GoldMine galaxies have no measured HI radius, we now apply to all
  galaxies the assumptions described in Sect.\ref{s:obsass} about the initial HI 
  profile and deduce the current HI radius from present mass and deficiency 
  by Eq.\ref{e:massratio}. Then Eq.\ref{e:pmax} gives the estimate for the 
  {\it required ram pressure} needed to produce the observed deficiency. 

\begin{figure}
  \includegraphics[width=0.33\textwidth, angle=270]{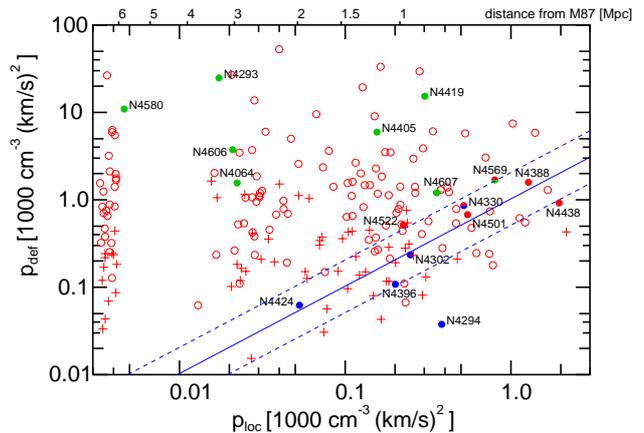}
  \caption{
    Similar to Fig.\ref{f:PlocPcfg}, but for the ram pressures computed from the 
    deficiency and assuming that the HI disk follows a Miyamoto-Nagai profile 
    with radial scale equal to the optical radius. This comprises the entire sample 
    taken from GoldMine. The group of objects at very low local ram pressure are
    galaxies of the 23 Mpc cloud, computed under the assumption if the ICM model 
    for the main cluster were extrapolated to that distance.}
  \label{f:PlocPdefR}
\end{figure}
 
  In Fig.\ref{f:PlocPdefR} we compare the required ram pressures with the local ram 
  pressures. The overall picture is very similar to Fig.\ref{f:PlocPcfg}. The required 
  and maximum ram pressures in the subsample of objects with VIVA data agree within a 
  factor of 2. Non-deficient galaxies are absent, because their negative deficiencies 
  would require negative ram pressures. These include NGC~4254 despite its long HI tail. 
  This leaves 197 galaxies. 

  \citet{vollmer01} and subsequent modelling work treat the ISM as a population of
  HI clouds instead of a smooth gas disk. Appendix \ref{s:vollmer} shows that an
  analytical treatment of this approach yields very similar results, albeit giving 
  systematically higher required ram pressures which are somewhat below the pressures 
  used in the individual modellings by Vollmer.    

\subsubsection{Catalog of Galaxies with Active and Past Stripping}

  These results allow us to identify in the GoldMine sample all objects near and below the 
  line of equality as candidates for undergoing stripping at the present time, while those 
  well above the line with $p_{\rm def} > 2 p_{\rm loc}$ are candidates that were stripped 
  in the past. Since the accuracy of the ram pressures estimates will not be better 
  than about a factor of 2, objects near the line of equality can be of either type. 
  Choosing the upper line in the figure as the border between the types gives some 
  preference to pick galaxies actively subject to stripping. 
  Tables \ref{t:cannow} to \ref{t:canpast3} list these two groups of objects, whose 
  inclination-corrected rotational velocities are above $30\speed$. Among the 50 galaxies 
  with active stripping (also including NGC~4522) are 23 low deficiency objects ($def < 0.3$);
  147 objects must have had stripping in the past, including 16 low deficiency galaxies. 
  A further 35 non-deficient galaxies ($def \le 0$; not displayed in Fig.\ref{f:PlocPdefR};
  Table \ref{t:cannodef}) complete our sample.  

  Except for the 39 objects in common with VIVA, the majority of the entire sample lacks
  HI maps, which is indicated in the tables as the absence of an entry for the isophotal
  HI radius from VIVA. Thus, among high deficiency objects in Table \ref{t:cannow} one finds
  19 candidates with active stripping: NGCs 4413, 4438, and 4540, ICs 797, 3059, 3142, 3105, 
  3239, 3258, 3311, 3365, 3414, 3453, 3476, 3583, and 3611, and VCCs 328, 1605, and 1644. 
  They are rather small with optical diameters between 1 and 3 arcmin, and are between 1 and $6\deg$
  of M87. High-resolution HI maps of these galaxies might very likely show direct signs for 
  the ICM-ISM interaction, such as HI disks with extensions and asymmetries. Likewise, one 
  could expect that the 85 high-deficient objects with past stripping would show well-truncated 
  HI disks. 

\subsubsection{Interpretation of the $p_{\rm loc} - p_{\rm cfg}$ Diagram}  

  In the idealized case of the long-pulse limit, a galaxy has always time to fully respond 
  to the external ram pressure $p_{\rm loc}$, given by the density of the ICM at the 
  distance from the cluster centre and the galaxy's speed. Therefore the HI disk will
  always be stripped to the stripping radius, where the external ram pressure is balanced
  by the galaxy's restoring force. Thus the maximum ram pressure $p_{\rm cfg}$ estimated
  from the outer rim of the HI disk is equal to the external ram pressure.

  During the infall of a galaxy into the cluster the external ram pressure $p_{\rm loc}$ 
  first increases with time, then reaches a peak value at the closest approach to the cluster
  centre, and finally decreases again. Consequently, $p_{\rm cfg}$ first follows $p_{\rm loc}$
  closely, up to the moment of peak pressure at pericentric distance. After that $p_{\rm loc}$
  decreases, while $p_{\rm cfg}$ remains constant as there will be no further stripping. 
  This produces a very simple evolutionary track in the $(p_{\rm loc}, p_{\rm cfg}$ diagram: 
  The first part follows the line of equality up to the position of peak pressure, and is
  independent of the orbital trajectory and the properties of the galaxy. The second part 
  at constant $p_{\rm cfg}$ is determined by the peak pressure at pericentric distance. 

\begin{figure}
  \includegraphics[width=0.33\textwidth, angle=270]{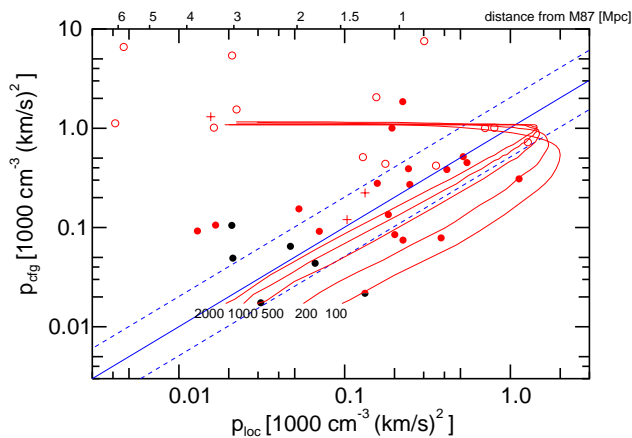}
  \caption{
    Like Fig.\ref{f:PlocPcfg}, but red dots mark galaxies which show extra-planar gas, 
    one-sided extension of the disk, or asymmetric morphology in VIVA's HI maps. Black 
    dots indicate non-deficient galaxies. For the mass model of NGC~4522 \citep{vollmer06} 
    evolutionary tracks are computed from test particle models with Lorentzian pulses 
    of various FWHM durations (as indicated in Myr). The track for the longest pulse has 
    a maximum ram pressure of 1400~cm$^{-3}$(km/s)$^{2}$; for shorter pulses the pressure 
    is raised to give the same stripped mass fraction of about 0.9, i.e. $def=1$.}
  \label{f:PlocPcfgTails}
\end{figure}

  Since such an evolutionary track is determined only by the ram pressure history, it can 
  be computed by test particle models (as in Section \ref{s:ring}) where a galaxy 
  is subjected to a ram pressure pulse. The tracks in Fig.\ref{f:PlocPcfgTails} are computed 
  for the mass model of NGC 4522 \citep{vollmer06} and pressure pulses with Lorentz profile 
  of various durations. While the conditions at pericentric distance determine the maximum 
  ram pressure, the orbit's eccentricity determines the pulse duration. The peak pressure 
  for the longest pulses is $1400\press$, which corresponds to a pericentric distance of 
  about 0.35 Mpc. The galaxy's gravitational restoring force also includes the enhancement 
  factor due to the molecular phase. 

  In Fig.\ref{f:PlocPcfgTails} the VIVA sample is shown in a plot similar to Fig.\ref{f:PlocPcfg}
  but emphasizing the galaxies with HI tails, extraplanar gas, or any clear asymmetry in their
  HI distribution. The figure shows that for pulse durations longer than 1000 Myr -- which 
  amounts to 5 periods for vertical oscillations in the outer parts of the galaxy 
  (cf. Fig.\ref{f:vertperiod}) -- the tracks follow the line of equality and show a sharp turn 
  into the horizonal part at peak ram pressure. As they are within a factor of 2 of the equality 
  $p_{\rm cfg} = p_{\rm loc}$, the maximum pressure $p_{\rm cfg}$ deduced from the stripping 
  radius is indeed a good measure of the actual ram pressure. The break point between diagonal 
  and horizontal tracks does not coincide exactly with the condition $p_{\rm cfg} = p_{\rm loc}$. 
  This is due to deviations of the rotation curve of the model for NGC~4522 from a constant speed, 
  which is supposed in the computation of $p_{\rm cfg}$ and the radial dependence of the vertical 
  oscillation period. 

  Shorter pulses give substantial deviations from the equal pressure line. At typical pulse 
  widths of 300 Myr (cf. Fig.\ref{f:clusters}, and as found by Vollmer from detailled modellings)
  the inferred maximum pressure $p_{\rm cfg}$ tends to underestimate the true ram pressure at 
  that instant. Furthermore, at the time of peak pressure the deduced maximum pressure still 
  remains below its terminal value: in short-pulse stripping events a substantial amount of the 
  gas removal occurs only {\it after} the peak ram pressure, as gas parcels are first pushed 
  away from the galactic plane but then require more work and time to escape.

  Given the diverse uncertainties involved in the ram pressure estimates, discussed in 
  Section \ref{s:uncertain}, it is quite remarkable how well the galaxies of different HI 
  morphologies occupy different regions in this diagram: objects with signatures of 
  ongoing stripping such as HI extensions are close to the line of equality, and 
  galaxies with well-truncated disks tend to have large distances from the cluster 
  centre. Non-deficient galaxies are found near the line of quality, with low inferred
  ram pressures and at large distances from the centre. This suggests that they 
  could be galaxies at the start of their infall into the cluster, during which
  they become active strippers, and after passage close to the centre to become
  highly deficient objects with well-truncated HI disk and to emerge again to large
  distance from the centre. This is consistent with the findings by \citet{yoon17}.

\section{Conclusions}\label{s:concl}

  Ram pressure stripping is thought to be the dominant gas removal mechanism for galaxies 
  in clusters. A full treatment of this process requires hydrodynamic numerical simulations 
  employing complex and detailed physics (such as heating from the ICM, gas cooling by 
  various processes, and stellar feedback). The often used criterion by \citet{gunngott72} 
  provides a simple analytic approximation from which the amount of stripped gas can be 
  estimated. We have extended this quasi-static criterion to ram pressure pulses of 
  arbitrary duration. 
  
  A galaxy falling into a cluster moves on a highly eccentric trajectory through the 
  ICM and experiences a ram pressure which increases as it approaches the inner cluster 
  with the denser ICM. At the closest distance to the cluster centre the ram pressure 
  reaches a peak value, and on the way out the pressure decreases again. Thus, 
  the ram pressure is a pulse (of e.g. Lorentz profile) characterized by a maximum value 
  and duration. 
  We consider the motion of gas parcels in the galaxy's gravitational field and under 
  the influence of this pressure pulse. We find that the outcome of a stripping event 
  -- the amount of stripped gas and the stripping radius -- is determined by the duration 
  of the pressure pulse in relation to the period for oscillations perpendicular 
  to the plane:
  \begin{itemize}
     \item{} with long pulses one approaches the criterion of \citet{gunngott72}, 
             where the outcome depends on the maximum ram pressure $p_{\rm max}$. 
     \item{} for short pulses the outcome depends on the time-integrated ram pressure  
             $(v\Sigma)_{\rm ICM}$, because the acceleration of gas parcels 
             is governed by momentum transfer.
  \end{itemize}
  With these two limiting cases a comprehensive analytic description is provided 
  for the results of stripping due to pressure pulses of any duration and shape.
  For a galaxy, the amount of stripped gas can be estimated from the ram pressure
  history and vice versa.   

  Simple numerical models, based on this kinematical treatment of the motion of gas 
  parcels in the gravitational field of a galaxy, give a very good agreement with the 
  results of the SPH simulations of \citet{jachym07, jachym09}, which treat the dynamics 
  of the gas more accurately. Thus it is possible to reliably estimate for the face-on 
  stripping situation how much gas is removed from the galaxy for good and also how much 
  is only pushed out of the plane but will eventually fall back into the disk. Likewise, 
  the radii beyond which gas escapes or is only pushed away from the plane can be predicted. 
  That purely kinematic arguments suffice to provide a good approximation indicates that
  for these quantities more intricate details of fluid dynamics do not play a major role.

  In normal galaxies the periods for vertical oscillations about the galactic plane
  range between 50 and 200 Myr. This restricts the validity for the long-pulse limit, 
  viz. the criterion of \cite{gunngott72} to stripping events that are substantially 
  longer than these values. If the pulse duration is so short as to approach the period 
  for vertical oscillations, the criterion is no longer strictly valid and tends to 
  underestimate the maximum ram pressure needed for a given deficiency. This is
  an important caveat which must be kept in mind when using the criterion 
  of \citet{gunngott72}.

  Cosmological N-body simulations suggest that the most likely orbits of infalling
  satellites are highly eccentric and reach pericentric distances of 20 percent of the 
  virial radius. From the distributions of ICM and dark matter in clusters, deduced 
  from X-ray observations, we find that in a wide range of clusters the typical pulse 
  widths are concentrated around about 300 Myr. This implies that 
  \begin{itemize}
    \item In massive clusters the long-pulse limit is the more appropriate 
          approximation for the interpretation of stripping events. Deficiency and
          stripping radius are determined by the peak ram pressure encountered 
          by a galaxy.
    \item In less massive clusters and groups the transfer of momentum becomes 
          increasingly important. Thus the application of the quasi-static criterion 
          of \citet{gunngott72} will tend to underestimate the necessary ram pressures.
  \end{itemize}
  Together with reasonable approximations for spiral galaxies (flat rotation curve, 
  initial parameters of the HI gas disk, centrifugal acceleration as proxy for maximum
  restoring force, correction for the presence of molecular gas) the long-pulse limit 
  can be used to analyse observations of galaxies in large samples with limited knowledge 
  of their gas content or surrounding ICM conditions. Only a very few observational 
  parameters are required: present HI mass, HI deficiency, optical diameter, and 
  rotational velocity. Application of this analytic method to the deficient galaxies 
  of the Virgo cluster shows that in galaxies with clear signs for ongoing stripping the 
  ram pressure required for the measured deficiencies matches well with the value 
  estimated from the ICM density and the estimated flight speed. The comparison of 
  these two pressure estimates then allows us to identify active and past strippers 
  among galaxies whose stripping phase had not yet been determined. 
  Our analysis has identified 19 highly HI-deficient galaxies not previously 
  mapped by VIVA which are clear candidates for active stripping. Interferometric 
  HI imaging could reveal direct signs of the ongoing ICM-ISM interaction. Furthermore, 
  in 85 past strippers one expects clearly truncated HI disks.

  
\section*{Acknowledgements}

Hearty thanks go to Richard W\"unsch who asked a question which turned
out to be the clue to the complete solution.
This research has made use of the GOLDMine Database operated by the Universita' 
degli Studi di Milano-Bicocca. We gratefully acknowledge support by the 
grant project 15-06012S of the Czech Science Foundation, the project LM2015067
of the Ministry of Education, Youth and Sports of the Czech Republic,
the institutional research project RVO:67985815., the Albert Einstein Center for
Gravitation and Astrophysics via the Czech Science Foundation Project 14-37086G.
We thank Edmunds Optics whose catalogue slogan provided an inspiration for the title.




\appendix{}

\section{Long pulse limit}
  \label{s:long}  
  The basic response of a gas element to ram pressure is well illustrated by a simple model:
  A particle initially at rest at the centre of the potential $\Phi(z) = -0.5/(1 + z^2)$, 
  with the height $z$ above the galactic plane, is subjected to a pulse of external force. 
  Using scaled units in which the escape speed from the centre is unity, the period for 
  oscillations about the rest position is $2\pi$. The maximum restoring force occurs at 
  $z=1/\sqrt{3} = 0.577$ and is 0.3247 in these units. Since the potential at that point 
  is $-3/8$, the particle still remains gravitationally bound. In order to escape it must 
  thus either have already sufficient speed at this height or need further acceleration by 
  a continuing external force.

\begin{figure}
  \includegraphics[width=0.33\textwidth, angle=270]{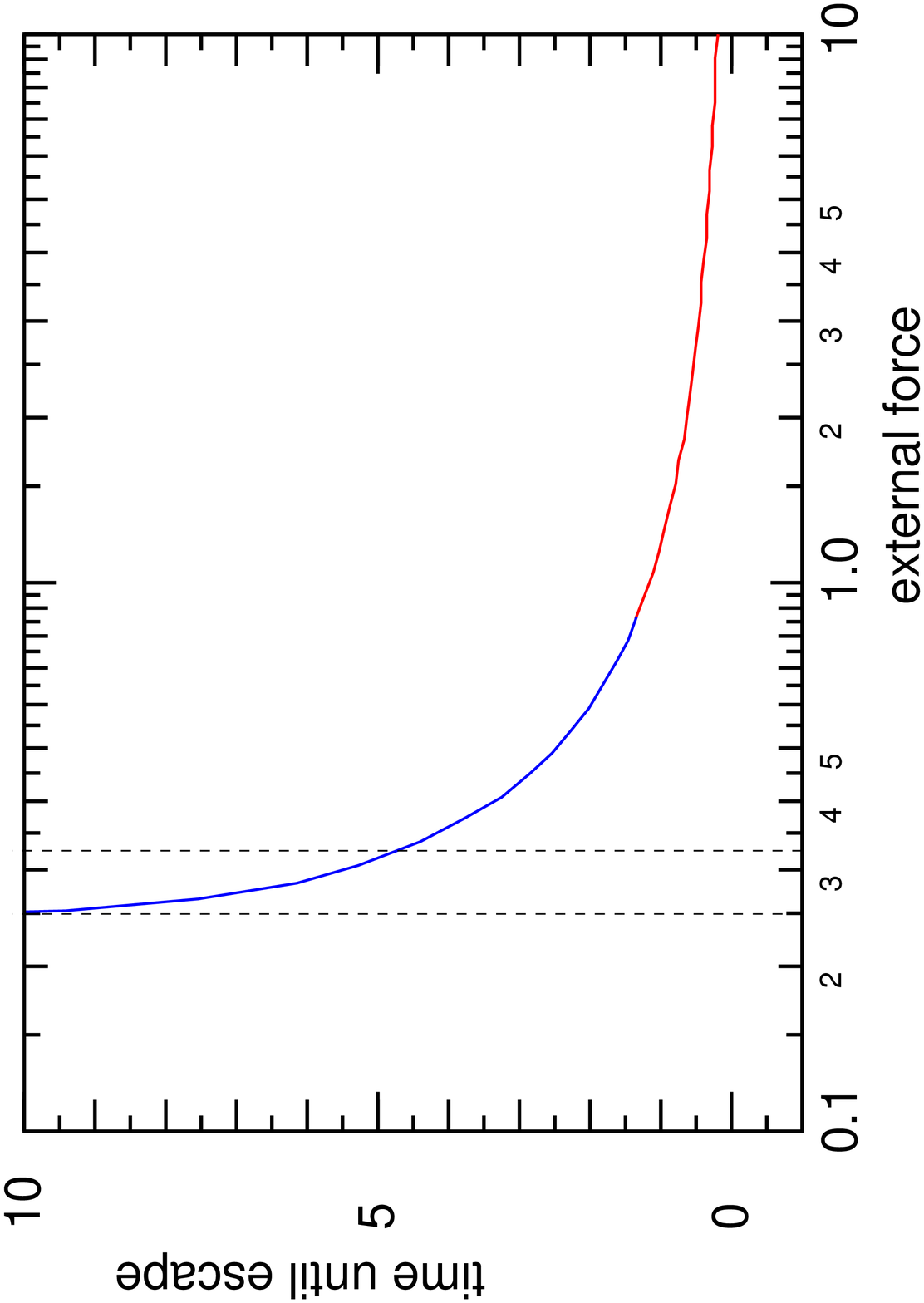}
  \caption{
    The times for a particle to escape from the potential well as a function of the external
    force, which rises at time zero rapidly like $(1-\exp(-t/\tau))$ with $\tau = 0.1$
    to a constant value. In the red part of the curve the particle attains positive 
    total energy before reaching the height of maximum restoring force, in the blue part 
    it is accelerated to escape speed only after passing that height. The right vertical
    dashed line indicates the maximum restoring force, the left vertical one marks the 
    minimum value for a constant force necessary to cause escape.}
  \label{f:escapes}
\end{figure}

  For a constant force, Fig.\ref{f:escapes} shows that the time until the particle escapes 
  decreases with higher force: For a strong force (more than about thrice the maximum restoring 
  force) the particle's total energy becomes positive even before the particle reaches the 
  height of maximum restoring force. It thus is quickly dislodged. If the force is less than 
  this value, the particle needs some time -- of the order of a quarter period of vertical 
  oscillations: about 1.5 time units -- to reach the height for maximum restoring force. 
  Thereafter it receives the acceleration to escape speed by the continuing external force, 
  for which a smaller but longer force pulse is sufficient. 

  The galaxy's orbit and the ICM distribution determine the height and duration of the 
  ram pressure pulse. Using various pulse forms different galaxies exhibit a similar general 
  behaviour to that depicted in Fig.\ref{f:escapes}. The minimum ram pressure necessary to 
  dislodge a gas parcel is influenced somewhat by the shape and duration of the ram pressure 
  pulse: for this simple model the quasi-static estimate is $F_{\rm min} = 0.3247$, with a 
  constant force it can be shown to be $0.25$, for Gaussian and Lorentzian pulses one gets 
  0.35 and higher values for shorter pulses. But as these values are quite close, one may 
  well use the more convenient quasi-static formulation.
  
\section{Efficiency of Momentum Transfer}
  \label{s:efficiency}

  If the pressure pulse is much shorter than the period of vertical oscillation, the gas element acquires 
  the full momentum $(v\Sigma)_{\rm ICM}$ imparted to it, and behaves accordingly, as described by 
  the short-pulse limit. What happens for longer pulse durations can be understood in the mechanical 
  analogy of a spring balance: When it is compressed by a slowly increasing external force -- this 
  corresponds to the slow increase of the ram pressure during the galaxy's flight towards the cluster 
  centre -- its deflection from the resting position will faithfully reproduce the static 
  force-deflection relation $F(z) = -D z$. This pertains as long as the force does not exceed some 
  limit value for its elasticity, which corresponds to the maximum value of the restoring force in 
  a galactic disk. When the force slowly subsides, the balance will return to its rest position, again 
  following the force-deflection relation. Thus, a slow and long pulse will not have any effect apart 
  from pushing temporarily the gas away from its rest position. However, the momentum provided 
  ($\int F(t)dt$) does not include the work done against the spring during compression 
  or obtained from it during release:
\begin{eqnarray}
   F(t)         &=& m \dot{v}  - Dz \\
   \int F(t) dt &=& m\Delta v - D \int z(t) dt 
\end{eqnarray}
  Thus, the ratio of the momentum deposited $m\Delta v$ and the momentum provided ($\int F(t)dt$) 
  describes to what extent momentum transfer is involved in the interaction. Let us call this ratio 
  {\it momentum efficiency}. 

\begin{figure}
  \includegraphics[width=0.33\textwidth, angle=270]{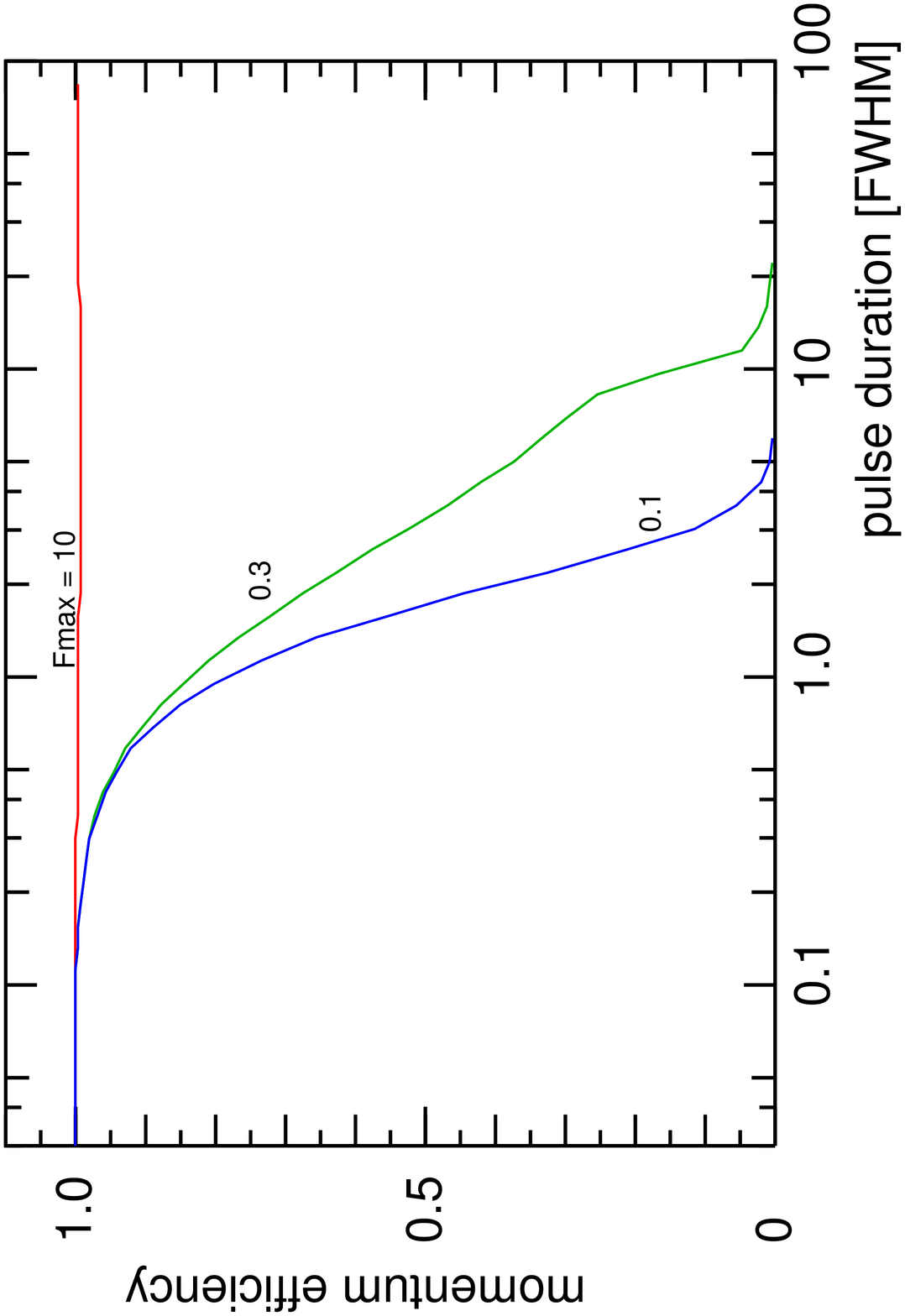}
  \caption{
    The fraction of the momentum deposited on a test particle in a schematic potential 
    well which is subjected to a Gaussian force pulse with peak value $F_{\rm max}$ and 
    various FWHM durations. The minimum force to overcome the well's restoring force is 
    0.3247, and the period for small oscillations about the rest position is $2\pi$, in 
    the normalized units of the model. The red curve is for cases where the particle escapes 
    from the well, for the green and blue curves the particle remains bound, but is perturbed 
    by the force pulse to oscillate about the rest position.}
  \label{f:efficiency}
\end{figure}

  The dependence of this quantity on the duration of the pressure pulse is summarized in 
  Fig.\ref{f:efficiency}, for the simplified model (as in Sect.\ref{s:long}) of a particle 
  initially at rest at the bottom of a potential $\Phi(z) = -0.5/(1 + z^2)$, then 
  subjected to a Gaussian force pulse with FWHM duration $\delta$ and maximum height 
  $(1+\delta)/\delta$. The simulation covers the time from $3\delta$ before the peak force 
  to $3\delta$ after it. 

  If the pulse is strong enough to make the particle escape, all the provided momentum is deposited 
  in the particle, independent of pulse duration or shape.
 
  For weaker pulses the particle remains in the potential well, but the outcome depends on the duration
  of the pulse: In short pulses all the momentum is deposited in the particle, which then oscillates about 
  the rest position with a period of $2\pi$ in the scaled units of the model and an amplitude corresponding
  to the deposited momentum. If the pulse is longer than this characteristic period, the particle 
  is merely displaced from the rest position and then brought back, without any momentum remaining. 
  The green curve in Fig.\ref{f:efficiency} shows that if the peak force is 0.3, just below the minimum force 
  to leave the galaxy, the momentum transfer becomes somewhat enhanced, which pushes the particle into 
  oscillations that approach the escape conditions.   

  This behaviour is characteristic for the test particle models for galaxies, which are presented in 
  Sect.\ref{s:ring}: While the stripped gas elements always absorb all the available momentum, gas 
  parcels that would remain gravitationally bound receive their full share only for very short pulses, 
  in accordance with the short pulse limit. At longer pulses, they receive a progressively lower fraction. 
  For pulse durations longer than about 200 Myr for a Milky Way type galaxy the momentum efficiency for all 
  bound gas parcels nearly vanishes. This also reduces the fraction of gas which is kicked out of the disk 
  and would later be reaccreted. Thus, in long pulses the gas disk is divided in an outer part escaping and 
  an inner part remaining near the disk plane, without any intermediate part of reaccretable gas.  

\section{Parameters of HI distribution in galaxies}
  \label{s:params}
   
  The initial distribution of the gas in a galaxy is a very important parameter that
  influences how much gas can be removed from a galaxy subjected to a given ram pressure.
  It is impracticable or impossible to infer on the initial state of the gas disk in 
  individual galaxies in our sample, which differ substantially both in their present 
  gas disk and their basic parameters, such as size, rotation speed, etc. However, 
  we can use these data to deduce which general properties the unperturbed gas disks 
  must have had. This Section shows that reasonable assumptions of the HI radial scale 
  and the ratio of initial HI radius and optical radius lead to satisfactory reproduction
  of the data.
  
\begin{figure}
  \includegraphics[width=0.33\textwidth, angle=270]{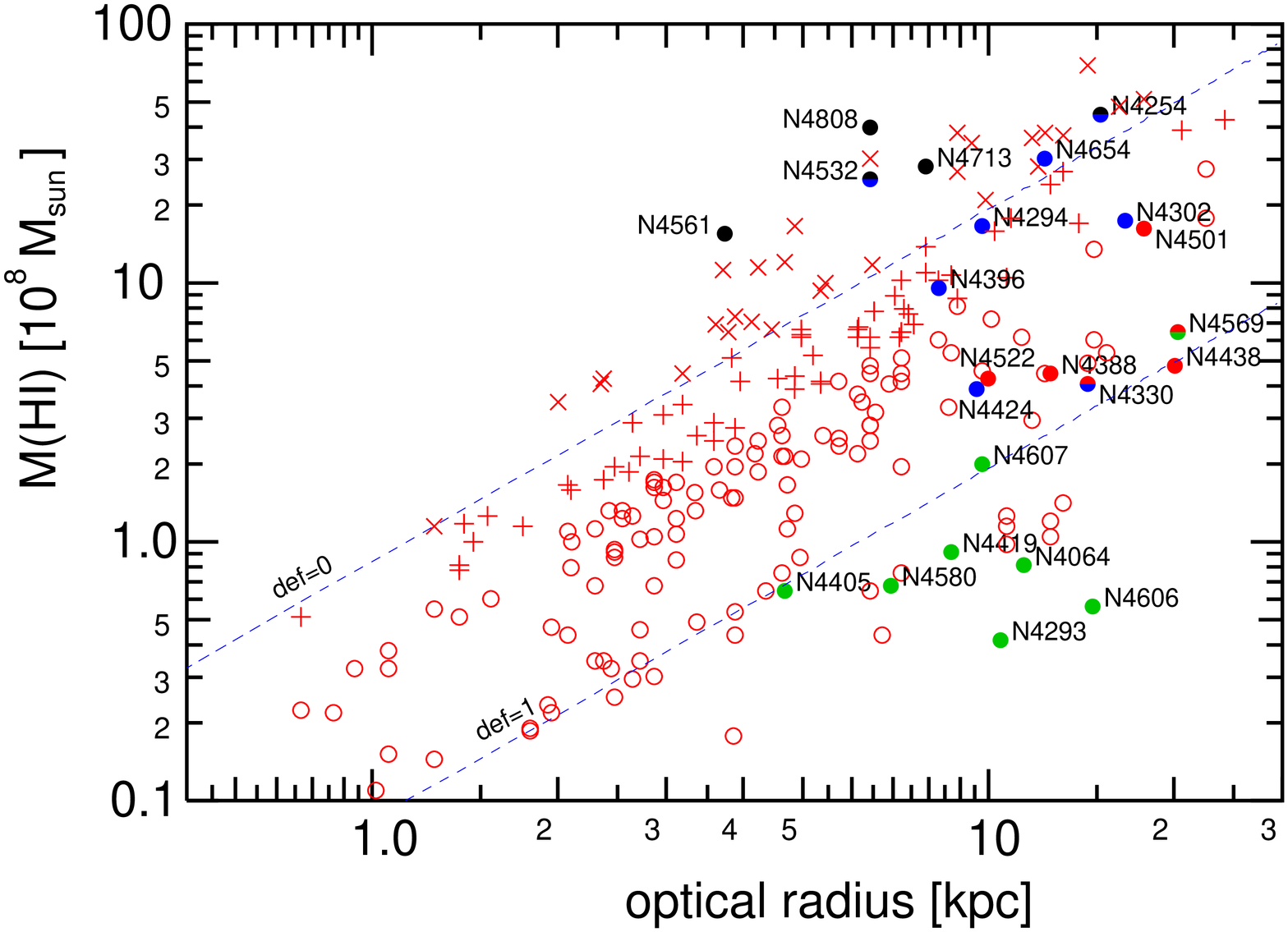}
  \caption{
    HI masses of galaxies in the Virgo cluster as a function of the
    optical radius (data from GoldMine). Black dots indicate several 
    examples for non-deficient galaxies, red dots mark the sequence
    of stripped objects \citep{vollmer09}, blue dots are galaxies
    with one-sided HI tails \citep{chung07}, and green dots are galaxies with
    clearly truncated HI disks, as found by inspection of HI maps in VIVA.
    Open circles are HI deficient galaxies ($def > 0.3$), plus signs mark 
    low-deficiency objects ($def < 0.3$), and X-signs are
    non-deficient objects  ($def < 0$), based on the formula of \citet{gavazzi13} 
    for deficiency. Two blue lines indicate the deficiencies of 0 and 1.}
  \label{f:RoptMhi}
\end{figure}

  The galaxies in the entire sample are shown in Fig.\ref{f:RoptMhi} as a relation between
  HI mass and optical radius. Among the galaxies with radii larger than about 5~kpc, 
  there are several objects that have been studied in detail. The HI maps from VIVA allow 
  us to distinguish different morphological features, such as tails and extensions of the 
  HI disk which signify ongoing stripping, and well-truncated HI disks which may be taken 
  as characteristic for stripping in the past. The galaxies of the sequence of \cite{vollmer09}, 
  which appear to be at maximum ram pressure or quite close to it, have loci near the line for 
  deficiency 1. Galaxies with one-sided tails from the HI disk \citep{chung07} are positioned 
  more closely to the line which corresponds to zero deficiency, thus indicating that they 
  are at an early stage of stripping. Objects whose HI maps (from VIVA) show a 
  well-truncated HI disk without any extending plume are found to be highly deficient. 
  NGC~4388 and NGC~4569
  also belong to this class. Galaxies whose HI diameter is larger than the 
  optical diameter, like normal spirals, have negative deficiencies. In the figure
  we mark and label only a few representative objects. NGC~4254 and NGC~4532 are
  non-deficient, but also have extensions of the HI disk.
  
  Most objects smaller than about 5~kpc are irregular galaxies, for which more
  detailed information are not available, such as HI maps. It is
  quite remarkable that they also follow the trend of the larger galaxies. 
  
  Figure \ref{f:RoptMhi} shows that the sample galaxies follow rather well
  a quadratic dependence of the HI mass and optical radius. VIVA adopts for 
  the formula to compute deficiencies a quadratic mass-diameter relation for 
  normal galaxies, with a corresponding constant HI surface density of 
  8.3~M$_\odot$/pc$^2$ (as averaged over the optical radius). In this paper 
  we use the formula by \citet{gavazzi13} 
  which also takes into account that the observed mass-diameter relation 
  deviates from a simple quadratic dependence:
  \begin{equation}
      def = 7.51-\log_{10}(M_{\rm HI}) + 0.68\times\log_{10}(d^2)
  \end{equation}
  with the optical diameter $d$ in kpc. The two relations 
  for zero deficiency coincide for an optical radius of about 6~kpc. 
  Thus, the choice of the formula to compute deficiencies is not overly critical.
  Both formulae give very similar results for normally sized galaxies; VIVA's formula 
  gives larger deficiencies at very large objects, and smaller values for dwarf galaxies.
 
  In most hydrodynamical and SPH simulations the gas in the disk is treated as 
  a continuous single-phase medium, whose surface density decreases outward.
  Our analytical considerations are also based on this assumption. Since the maximum
  restoring force decreases outwardly as well, for a given ram pressure the gas beyond 
  a certain `stripping radius' is removed. The increase of the ram pressure during the 
  flight of a galaxy towards the cluster interior thus causes a progressively stronger 
  truncation of the gas disk, but leaves the interior of the remaining disk unaffected. 
  Thus one expects that the remnant disk preserves the properties of the inner
  unperturbed galactic gas disk. 

\begin{figure}
  \includegraphics[width=0.33\textwidth, angle=270]{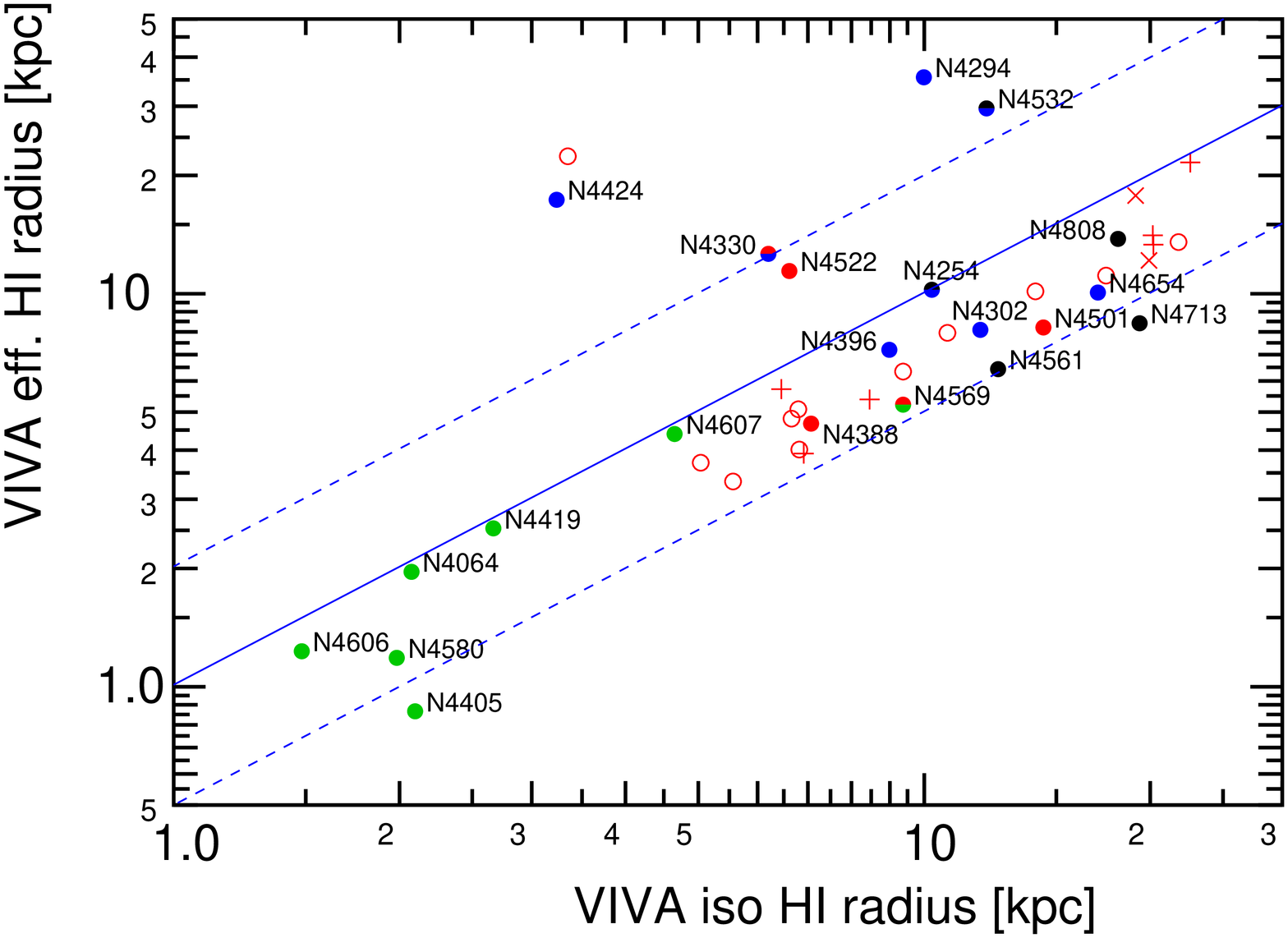}
  \caption{
    Comparison of the HI radii from VIVA: effective diameters $D^{\rm eff}_{\rm HI}$
    contain 50\% of the total flux and isophotal diameters $D^{\rm ISO}_{\rm HI}$ 
    where the azimuthally averaged HI surface density drops to $1\msunsigma$.
    Symbols are as in Fig.\ref{f:RoptMhi}. The full blue line indicates equality 
    of the radii, the dashed lines a factor of 2 in either sense.}
  \label{f:RisoReff}
\end{figure}

  VIVA gives two HI diameters: the isophotal one and the effective one, which 
  are compared in Fig.\ref{f:RisoReff}. The ratio of the two radii is an
  indicator for extraplanar HI or tails, as non-deficient galaxies have 
  $R_{\rm eff} \approx 0.7 \times R_{\rm ISO}$, but galaxies like NGC~4522 with 
  an extensive extraplanar HI plume have $R_{\rm eff} > R_{\rm ISO}$.
  The non-deficient NGC~4254 appears as a very odd object with 
  $R_{\rm eff} \approx 30 \times R_{\rm ISO}$, whose isophotal diameter is given 
  as an exceptionally low value of 0.15 arcmin, obviously a typo. We take the
  $R_{\rm eff}$ for the HI radius of this object. 

  The isophotal diameter uses a threshold of $1\msunsigma$ for the 
  surface density. As this value can be expected to give a reliable measure 
  for the extent of HI gas both in well truncated disks and normal galaxies 
  (cf. \citet{bigiel12}), we shall take the isophotal radius as the outer 
  HI radius, keeping in mind that objects with substantial extraplanar gas 
  can be misrepresented.

\begin{figure}
\centering
\includegraphics[width=0.33\textwidth, angle=270]{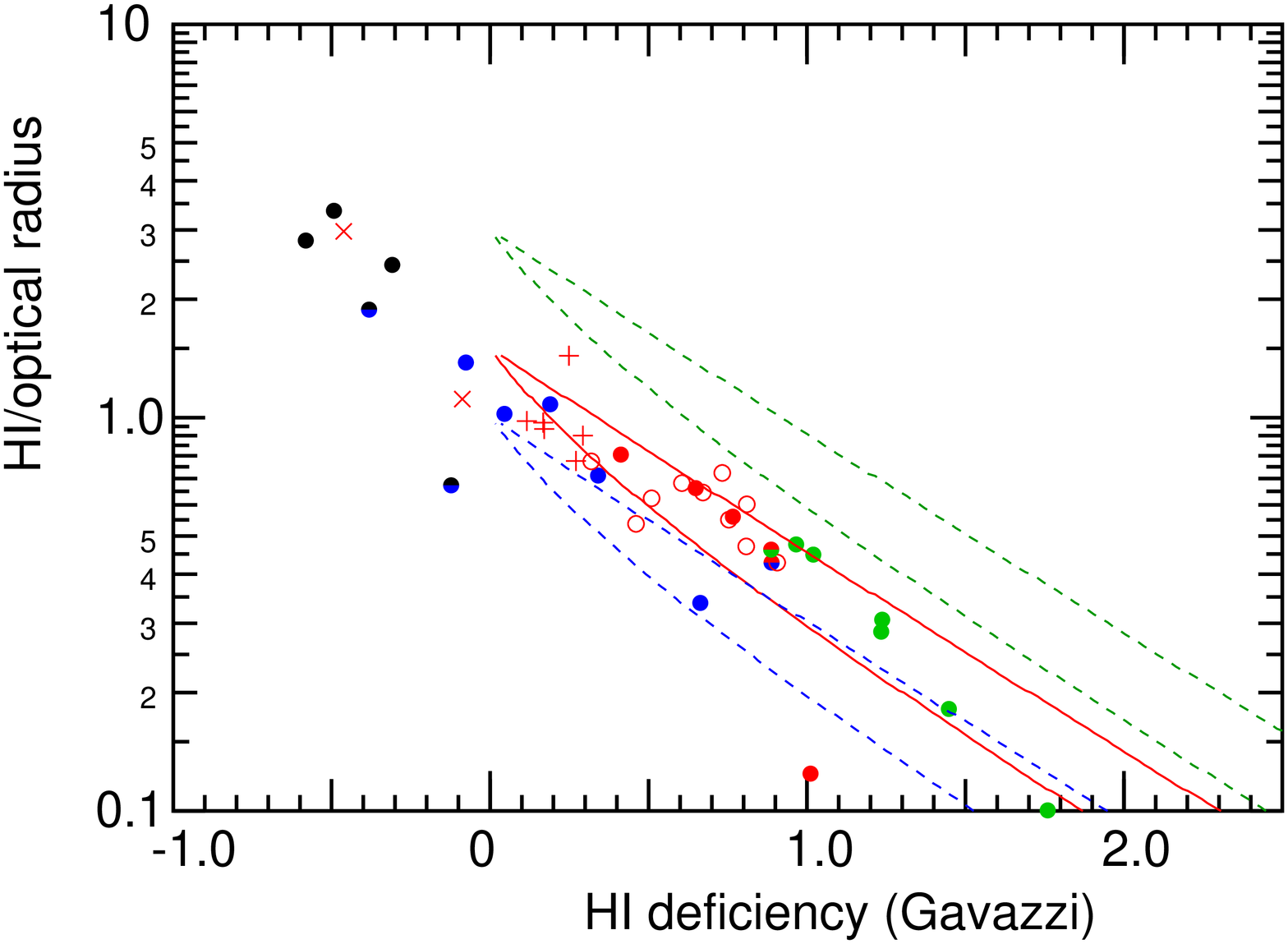}
\caption{ \small
  The ratio of the isophotal HI radius (from VIVA) and the optical radius (from
  GoldMine) of galaxies in the Virgo cluster as a function of HI deficiency 
  \citep{gavazzi13}. Symbols are as in Fig.\ref{f:RoptMhi}. The curves are the 
  expected relations for exponential HI disks with the ratio of radial scale 
  $R$ and initial HI outer radius $r_{\rm max}$ of the disk 
  $R/r_{\rm max} = 5$ and 0.5 (from top to bottom) and for the ratio 
  $r_{\rm max}/r_{\rm opt}$ of 1 (blue), 1.5 (red), and 3 (green).  
}
\label{f:defRRatio}
\end{figure}

  The dependence of the ratio of HI and optical radius on deficiency, as depicted 
  in Fig.\ref{f:defRRatio}, can be used to constrain the size of the initial HI 
  disk. The data are well reproduced, if one assumes that the HI radial scale is 
  close to the optical radius, and if the initial HI radius is about 1.5 times 
  the optical radius. This value agrees well with the results of 
  \cite{broeils97} from large undisturbed field galaxies which suggest 
  a value of 1.7. For our analyses we shall adopt that the initial HI disk
  is a Miyamoto-Nagai disk with radial scale equal to the optical radius,
  and is truncated at an outer radius 1.5 times the optical radius.

\begin{figure}
  \includegraphics[width=0.33\textwidth, angle=270]{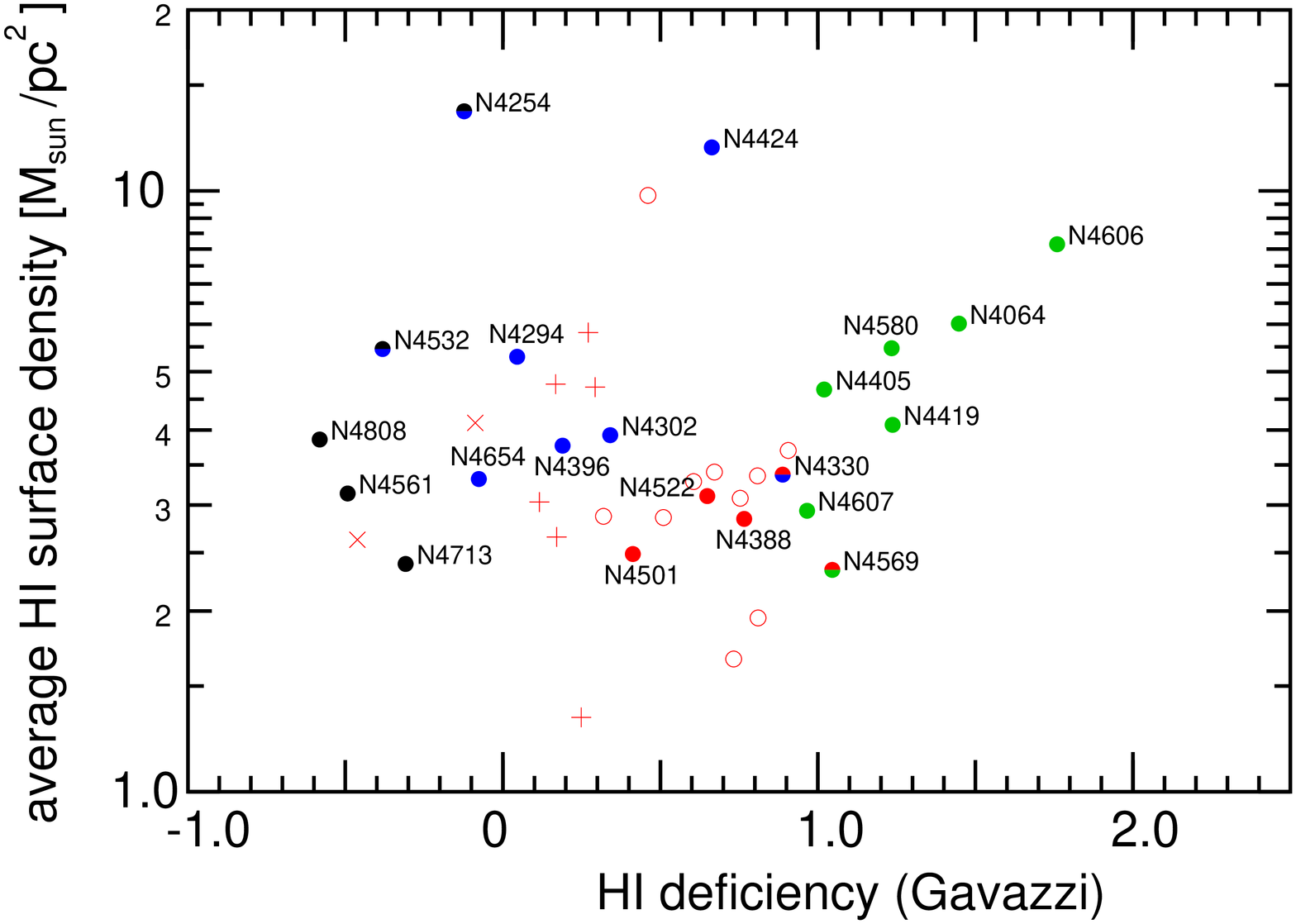}
  \caption{
    The mean HI surface density of galaxies in the Virgo cluster,
    computed from the measured HI mass (GoldMine) and the isophotal HI diameters 
    $D^{\rm ISO}_{\rm HI}$ (VIVA), within which the azimuthally averaged HI 
    surface density exceeds $1\msunsigma$, as a function of HI deficiency \citep{gavazzi13}. 
    Symbols are as in Fig.\ref{f:RoptMhi}.}
  \label{f:DefSigmaHI}
\end{figure}

  The mean HI surface density computed from the HI mass and the isophotal radius
  shows a remarkably small variation among the objects, as seen in Fig.\ref{f:DefSigmaHI}.
  Most galaxies have mean surface densities around $4\msunsigma$. 
  Although Fig.\ref{f:DefSigmaHI} does not show a strong dependence of the average
  HI surface density on deficiency, one notes that the highly deficient objects with 
  well-truncated HI disks have slightly higher surface densities than non-deficient 
  galaxies. As in both types of objects the isophotal diameters represent the true 
  extent of the HI gas, this difference in mean surface density yields another constraint 
  on the radial scale of the HI distribution: The mean surface density of a disk with 
  exponential profile, central density of $10\msunsigma$, and outer radius of 
  15 kpc is 5.3 or $3.9\msunsigma$ for radial scales of 15 and 10 kpc, respectively. 
  When truncated to deficiency 2, the mean surface density is close to the central value. 
  Thus, our assumption that the HI radial scale is equal to the optical radius reproduces 
  the weak trend seen in Fig.\ref{f:DefSigmaHI}. The initial central density would then 
  be taken from NGC~4606 as close to $8\msunsigma$. 

\section{Comparison with Vollmer's modellings}
  \label{s:vollmer}

  \citet{vollmer01} and subsequent works have modelled in detail a number of VIVA 
  sample galaxies. Since these objects serve here as a kind of calibration set, it 
  is worthwhile to ascertain to what extent our analytical approach is able to 
  reproduce also these models. The ISM was treated as a population of HI clouds 
  with different masses but fixed column density, and a sticky-particle algorithm was 
  used to model the interaction by inelastic collisions which give rise 
  to fragmentation, coalescence and mass exchange of clouds. Furthermore, to account 
  for clouds becoming denser closer to the galaxy's centre, an enhancement factor 
  was applied in the restoring force, as in Eq.\ref{e:pmax}.
 
  We compute from the HI deficiency the stripping radius, assuming a Miyamoto-Nagai
  disk, and then use Eq.\ref{e:pmax} with the fixed value for the gas surface density
  of $\Sigma_{\rm H} = 7.75\msunsigma$ and a fixed scale length of $R_0 = 2$ kpc, as 
  used by \citet{vollmer01} to compute the required ram pressure. This is done for
  models M, N, O, and P of \citet{vollmer01}, for which the actual gas density profiles 
  are not given, but we find that with a radial scale of 20 kpc for the Miyamoto-Nagai 
  gas disk we recover the maximum ram pressures by better than 6\%, and the stripping 
  radii within a factor 1.5.

\begin{figure}
  \includegraphics[width=0.33\textwidth, angle=270]{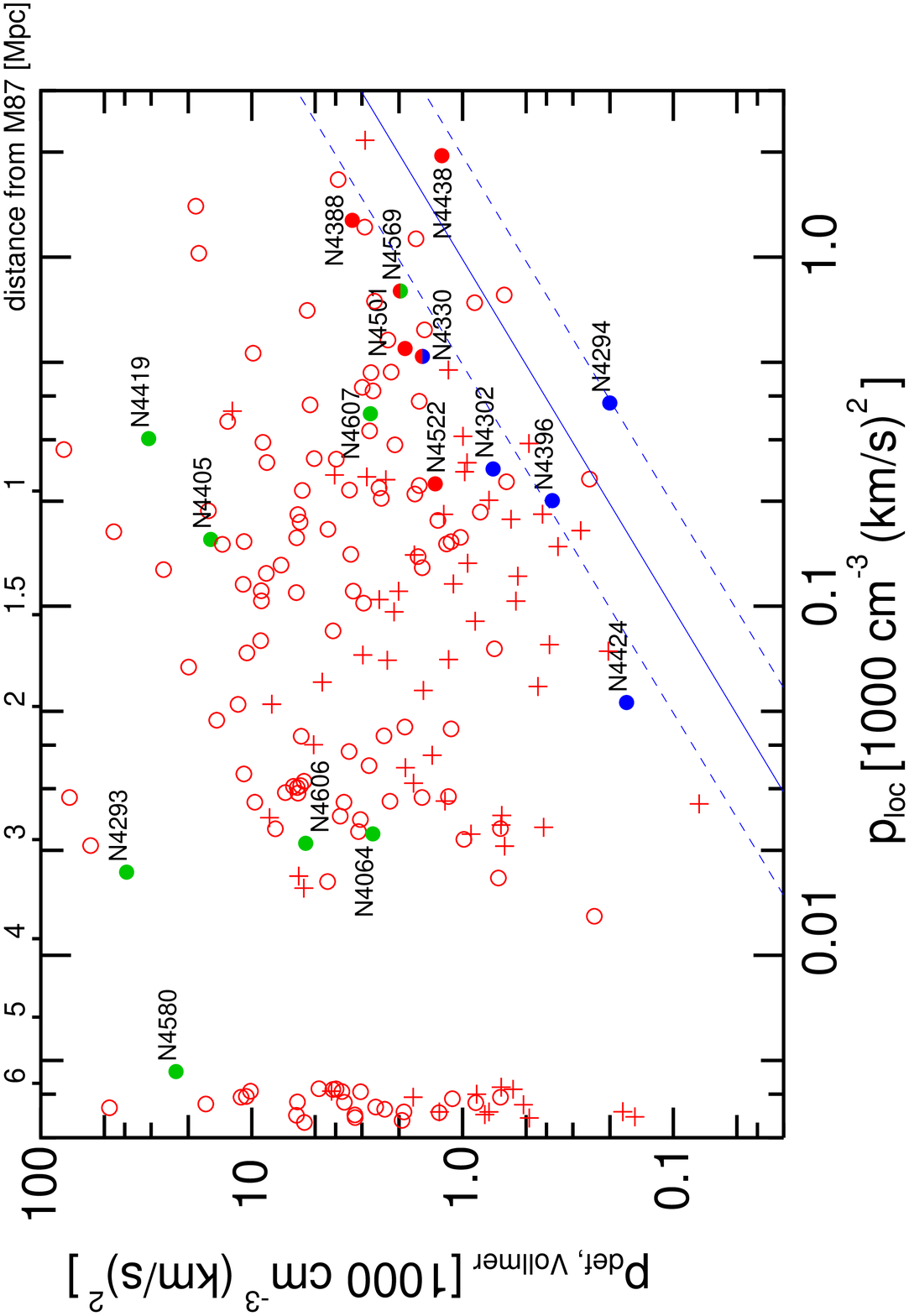}
  \caption{
    Similar to Fig.\ref{f:PlocPdefR}, but using approach of \citet{vollmer01}
    which represents the ISM as a population of HI clouds of equal column density.}
  \label{f:PlocPbernd}
\end{figure}

  With this prescription we now analyze the data of our full GoldMine sample, assuming 
  again that the HI disk has an initial outer radius of 1.5 times the optical radius and 
  follows a Miyamoto-Nagai profile with radial scale equal to the optical radius 
  (see Sect.\ref{s:params} for details). The results are shown in Fig.\ref{f:PlocPbernd}: 
  Because of the higher value for the HI column density for all objects, the required ram 
  pressures needed to account for the deficiencies are higher than in Fig.\ref{f:PlocPdefR}, 
  typically by about a factor of 3, which would be still acceptable in view of the
  overall uncertainties for the pressure estimates.  

  Although the required ram pressures are already higher than estimated from 
  Fig.\ref{f:PlocPdefR}, they still are systematically below the values used 
  by Vollmer in his individual modellings. For example, NGC~4522 needs $1350\press$ 
  (or 800 when $R_0$ is scaled with the optical radius (Eq.\ref{e:r0scaling})
  instead of 2000 \citep{vollmer06}. NGC~4388 requires 3300 (or 2600) instead 
  of 5000 \citep{vollmer03}, but \citet{damas16} report that a value three times lower is 
  needed, i.e. $1700\press$. Figure~5 of \citet{vollmer09} also shows that 
  for most of the galaxies the maximum ram pressure is higher (factor of 2 and more) 
  than estimated from the local ICM density. While this factor is still acceptable within the 
  overall uncertainties of the data, a possible explanation lies with the assumed gas 
  distribution. The modellings unfortunately do not provide information about the initial 
  or the current gas density profile, such as radial scale and initial outer radius. 
  Had we adopted radial scales substantially smaller (about one half) than the optical 
  radius, we would get indeed required ram pressures as high as the published values.

\section{The Catalogue}
\begin{table*}
  \centering
  \caption{Virgo galaxies which undergo stripping at present, with $p_{\rm def} < 2 p_{\rm loc}$,
     sorted in descending order of HI deficiency. The columns are HI deficiency, linear distance
     from M87 [Mpc],  optical radius [kpc], HI isophotal radius [kpc] (from VIVA, when available),
     rotational velocity [$\speed$], local ram pressure [$1000\press$], 
     maximum ram pressure estimated by the centrifugal acceleration at the observed HI radius 
     [$1000\press$], and the ram pressure needed to produce the current HI deficiency 
     [$1000\press$].}
  \label{t:cannow} 
  \begin{tabular}{lcccccccc}
    \hline
 Name & $def$ & $d$ & $r_{\rm opt}$ & $r_{\rm HI, iso}$ & $v_{\rm rot}$ & $p_{\rm loc}$ & $p_{\rm max}$ & $p_{\rm def}$ \\  
    \hline \noalign{\rule{0pt}{0.6ex}}
NGC4438 & 1.01 & 0.29 & 20.1 & ---  & 134.1 & 1.95 & ---  & 0.92 \\ 
IC3611 & 0.98 & 0.66 & 4.4 & ---  & 37.6 & 0.47 & ---  & 0.79 \\ 
NGC4330 & 0.89 & 0.63 & 14.5 & 6.2 & 121.8 & 0.52 & 0.32 & 0.86 \\ 
NGC4413 & 0.8 & 0.32 & 7.2 & ---  & 99.1 & 1.67 & ---  & 1.3 \\ 
NGC4388 & 0.77 & 0.37 & 12.6 & 7.1 & 182.3 & 1.27 & 0.59 & 1.59 \\ 
NGC4424 & 0.66 & 1.98 & 9.6 & 3.2 & 34.5 & 0.05 & 0.07 & 0.06 \\ 
NGC4522 & 0.65 & 0.98 & 10.0 & 6.6 & 105.7 & 0.22 & 1.85 & 0.52 \\ 
IC3476 & 0.63 & 0.51 & 6.4 & ---  & 91.1 & 0.75 & ---  & 0.74 \\ 
NGC4402 & 0.61 & 0.4 & 9.8 & 6.7 & 122.5 & 1.13 & 0.28 & 0.62 \\ 
NGC4540 & 0.57 & 0.98 & 6.4 & ---  & 78.3 & 0.22 & ---  & 0.44 \\ 
IC3258 & 0.57 & 0.51 & 6.4 & ---  & 57.8 & 0.74 & ---  & 0.24 \\ 
IC3239 & 0.56 & 0.59 & 3.1 & ---  & 45.6 & 0.58 & ---  & 0.48 \\ 
VCC1605 & 0.51 & 0.66 & 2.5 & ---  & 43.5 & 0.47 & ---  & 0.54 \\ 
VCC328 & 0.49 & 0.85 & 2.5 & ---  & 39.1 & 0.29 & ---  & 0.41 \\ 
IC3583 & 0.45 & 0.5 & 6.9 & ---  & 65.1 & 0.78 & ---  & 0.18 \\ 
IC797 & 0.42 & 0.81 & 4.2 & ---  & 80.5 & 0.32 & ---  & 0.57 \\ 
IC3365 & 0.42 & 1.07 & 6.1 & ---  & 68.0 & 0.19 & ---  & 0.21 \\ 
IC3059 & 0.41 & 1.19 & 4.6 & ---  & 60.0 & 0.15 & ---  & 0.26 \\ 
NGC4501 & 0.41 & 0.61 & 17.9 & 14.4 & 296.0 & 0.55 & 0.43 & 0.68 \\ 
IC3311 & 0.38 & 0.38 & 3.6 & ---  & 75.3 & 1.22 & ---  & 0.55 \\ 
IC3414 & 0.36 & 1.67 & 4.5 & ---  & 49.4 & 0.08 & ---  & 0.15 \\ 
IC3142 & 0.36 & 0.97 & 3.1 & ---  & 31.3 & 0.23 & ---  & 0.11 \\ 
IC3453 & 0.35 & 0.73 & 3.0 & ---  & 48.8 & 0.39 & ---  & 0.28 \\ 
NGC4302 & 0.34 & 0.93 & 16.7 & 11.9 & 187.6 & 0.25 & 0.25 & 0.23 \\ 
IC3105 & 0.34 & 0.96 & 6.4 & ---  & 46.1 & 0.23 & ---  & 0.07 \\ 
VCC1644 & 0.32 & 0.57 & 2.4 & ---  & 41.0 & 0.62 & ---  & 0.25 \\ 
NGC4321 & 0.32 & 1.17 & 22.5 & 17.5 & 268.8 & 0.16 & 0.27 & 0.27 \\ 
 \noalign{\rule{0pt}{0.6ex}}
NGC4595 & 0.29 & 1.08 & 5.3 & ---  & 90.4 & 0.18 & ---  & 0.29 \\ 
VCC512 & 0.28 & 0.66 & 3.6 & ---  & 56.5 & 0.47 & ---  & 0.21 \\ 
NGC4298 & 0.27 & 0.94 & 8.9 & 6.9 & 148.8 & 0.24 & 0.37 & 0.31 \\ 
VCC87 & 0.21 & 1.53 & 3.6 & ---  & 55.9 & 0.09 & ---  & 0.15 \\ 
NGC4633 & 0.2 & 1.03 & 6.1 & ---  & 99.9 & 0.2 & ---  & 0.19 \\ 
NGC4396 & 0.19 & 1.03 & 8.3 & 9.0 & 99.0 & 0.2 & 0.08 & 0.11 \\ 
NGC4470 & 0.18 & 1.36 & 4.5 & ---  & 88.5 & 0.12 & ---  & 0.23 \\ 
IC3446 & 0.18 & 0.27 & 2.7 & ---  & 80.9 & 2.16 & ---  & 0.43 \\ 
NGC4192 & 0.17 & 1.43 & 24.2 & 22.6 & 234.7 & 0.1 & 0.12 & 0.1 \\ 
NGC4222 & 0.17 & 1.08 & 8.7 & 8.5 & 113.0 & 0.18 & 0.13 & 0.12 \\ 
IC3562 & 0.16 & 0.83 & 2.1 & ---  & 38.2 & 0.31 & ---  & 0.13 \\ 
VCC1437 & 0.14 & 0.96 & 1.5 & ---  & 43.2 & 0.23 & ---  & 0.28 \\ 
NGC4498 & 0.12 & 1.33 & 7.0 & ---  & 116.1 & 0.12 & ---  & 0.14 \\ 
NGC4535 & 0.12 & 1.27 & 20.6 & 20.2 & 297.9 & 0.13 & 0.22 & 0.15 \\ 
NGC4639 & 0.1 & 0.91 & 7.9 & ---  & 176.3 & 0.26 & ---  & 0.23 \\ 
NGC4688 & 0.08 & 2.69 & 10.9 & ---  & 63.1 & 0.03 & ---  & 0.02 \\ 
IC3074 & 0.08 & 1.2 & 7.2 & ---  & 104.0 & 0.15 & ---  & 0.08 \\ 
VCC1726 & 0.07 & 1.65 & 3.2 & ---  & 45.0 & 0.08 & ---  & 0.06 \\ 
IC3742 & 0.07 & 1.1 & 5.0 & ---  & 95.4 & 0.18 & ---  & 0.12 \\ 
NGC4294 & 0.05 & 0.74 & 9.8 & 10.0 & 101.1 & 0.38 & 0.08 & 0.04 \\ 
NGC4206 & 0.04 & 1.14 & 12.6 & ---  & 137.5 & 0.16 & ---  & 0.04 \\ 
IC3591 & 0.0 & 1.68 & 3.8 & ---  & 49.9 & 0.07 & ---  & 0.03 \\ 
NGC4523 & 0.0 & 0.85 & 7.9 & ---  & 148.5 & 0.29 & ---  & 0.08 \\ 
    \hline
  \end{tabular}
\end{table*}

\begin{table*}
  \centering
  \caption{Like Table \ref{t:cannow}, but for galaxies which had undergone stripping in the past  
     (with $p_{\rm def} > 2 p_{\rm loc})$.}
  \label{t:canpast1} 
  \begin{tabular}{lcccccccc}
    \hline
 Name & $def$ & $d$ & $r_{\rm opt}$ & $r_{\rm HI, iso}$ & $v_{\rm rot}$ & $p_{\rm loc}$ & $p_{\rm max}$ & $p_{\rm def}$ \\  
    \hline \noalign{\rule{0pt}{0.6ex}}
NGC4606 & 1.76 & 3.03 & 14.8 & 1.5 & 76.7 & 0.02 & 0.67 & 3.75 \\ 
NGC4293 & 1.69 & 3.3 & 10.5 & ---  & 163.7 & 0.02 & ---  & 24.8 \\ 
IC796 & 1.47 & 1.2 & 3.9 & ---  & 56.6 & 0.15 & ---  & 9.02 \\ 
NGC4064 & 1.45 & 2.95 & 11.4 & 2.1 & 58.8 & 0.02 & 0.32 & 1.56 \\ 
NGC4309 & 1.4 & 6.42 & 6.7 & ---  & 164.3 & 0.0 & ---  & 26.54 \\ 
NGC4313 & 1.4 & 0.62 & 12.6 & ---  & 131.8 & 0.53 & ---  & 5.75 \\ 
NGC4312 & 1.34 & 1.11 & 12.6 & ---  & 109.6 & 0.17 & ---  & 3.51 \\ 
NGC4586 & 1.33 & 2.46 & 10.7 & ---  & 127.6 & 0.03 & ---  & 6.02 \\ 
VCC1623 & 1.32 & 1.3 & 1.8 & ---  & 54.4 & 0.13 & ---  & 19.37 \\ 
VCC1970 & 1.3 & 1.15 & 1.8 & ---  & 71.1 & 0.16 & ---  & 33.23 \\ 
NGC4307 & 1.29 & 6.21 & 13.2 & ---  & 162.5 & 0.0 & ---  & 6.29 \\ 
NGC4445 & 1.26 & 6.12 & 10.7 & ---  & 133.5 & 0.0 & ---  & 5.5 \\ 
NGC4419 & 1.24 & 0.84 & 8.7 & 2.7 & 195.7 & 0.3 & 3.01 & 15.3 \\ 
NGC4580 & 1.23 & 5.81 & 6.9 & 2.0 & 138.5 & 0.0 & 2.39 & 10.92 \\ 
NGC4356 & 1.22 & 6.23 & 10.7 & ---  & 147.5 & 0.0 & ---  & 5.95 \\ 
IC3718 & 1.21 & 1.01 & 6.4 & ---  & 36.2 & 0.21 & ---  & 0.79 \\ 
IC3392 & 1.21 & 0.79 & 7.2 & ---  & 111.0 & 0.34 & ---  & 6.08 \\ 
IC3374 & 1.08 & 6.08 & 3.9 & ---  & 42.2 & 0.0 & ---  & 1.69 \\ 
IC3077 & 1.06 & 1.23 & 2.9 & ---  & 31.9 & 0.14 & ---  & 1.48 \\ 
VCC517 & 1.05 & 2.26 & 2.5 & ---  & 170.2 & 0.04 & ---  & 52.68 \\ 
NGC4569 & 1.05 & 0.49 & 26.5 & 9.4 & 215.9 & 0.8 & 0.49 & 1.69 \\ 
IC3520 & 1.02 & 0.43 & 2.6 & ---  & 71.4 & 1.03 & ---  & 7.45 \\ 
NGC4405 & 1.02 & 1.17 & 4.7 & 2.1 & 102.3 & 0.16 & 1.35 & 5.95 \\ 
VCC513 & 1.0 & 3.05 & 1.8 & ---  & 102.5 & 0.02 & ---  & 26.84 \\ 
IC3412 & 0.99 & 6.08 & 3.9 & ---  & 50.3 & 0.0 & ---  & 1.77 \\ 
VCC1459 & 0.99 & 2.9 & 1.8 & ---  & 37.7 & 0.02 & ---  & 3.47 \\ 
VCC1784 & 0.97 & 1.14 & 2.0 & ---  & 30.0 & 0.17 & ---  & 1.85 \\ 
VCC1465 & 0.97 & 2.68 & 2.7 & ---  & 56.1 & 0.03 & ---  & 3.71 \\ 
NGC4607 & 0.97 & 0.77 & 9.8 & 4.7 & 91.5 & 0.36 & 0.3 & 1.21 \\ 
NGC4641 & 0.95 & 0.9 & 3.4 & ---  & 42.9 & 0.26 & ---  & 1.41 \\ 
IC800 & 0.94 & 0.91 & 4.6 & ---  & 80.6 & 0.26 & ---  & 2.96 \\ 
IC3716 & 0.94 & 1.63 & 1.9 & ---  & 44.3 & 0.08 & ---  & 3.63 \\ 
VCC1013 & 0.94 & 6.13 & 2.4 & ---  & 33.3 & 0.0 & ---  & 1.41 \\ 
IC3521 & 0.92 & 1.58 & 4.9 & ---  & 60.8 & 0.08 & ---  & 1.41 \\ 
NGC4380 & 0.91 & 6.1 & 11.8 & 5.0 & 151.3 & 0.0 & 0.7 & 1.98 \\ 
VCC1744 & 0.9 & 0.85 & 1.3 & ---  & 34.7 & 0.29 & ---  & 3.88 \\ 
VCC479 & 0.89 & 6.3 & 2.4 & ---  & 43.3 & 0.0 & ---  & 2.11 \\ 
VCC1821 & 0.89 & 1.77 & 1.0 & ---  & 46.0 & 0.07 & ---  & 9.54 \\ 
VCC1597 & 0.87 & 2.09 & 2.3 & ---  & 67.0 & 0.05 & ---  & 5.0 \\ 
IC3483 & 0.85 & 0.35 & 2.7 & ---  & 86.1 & 1.4 & ---  & 5.85 \\ 
NGC4579 & 0.81 & 0.53 & 15.6 & 9.4 & 278.6 & 0.7 & 0.86 & 3.05 \\ 
NGC4689 & 0.81 & 1.29 & 14.5 & 6.8 & 137.3 & 0.13 & 0.36 & 0.83 \\ 
VCC514 & 0.79 & 6.24 & 4.7 & ---  & 35.4 & 0.0 & ---  & 0.32 \\ 
VCC1574 & 0.78 & 0.87 & 1.1 & ---  & 101.8 & 0.28 & ---  & 29.53 \\ 
NGC4450 & 0.75 & 3.4 & 12.3 & 6.8 & 207.2 & 0.02 & 0.82 & 2.03 \\ 
NGC4287 & 0.74 & 6.66 & 4.9 & ---  & 86.0 & 0.0 & ---  & 1.55 \\ 
NGC4548 & 0.73 & 0.71 & 14.8 & 10.7 & 206.4 & 0.41 & 0.36 & 1.39 \\ 
IC3517 & 0.71 & 1.0 & 2.9 & ---  & 44.4 & 0.22 & ---  & 0.87 \\ 
VCC1208 & 0.71 & 2.61 & 2.1 & ---  & 49.5 & 0.03 & ---  & 1.86 \\ 
NGC4351 & 0.67 & 3.35 & 8.6 & 5.6 & 60.6 & 0.02 & 0.09 & 0.23 \\ 
VCC1931 & 0.66 & 0.9 & 3.1 & ---  & 60.7 & 0.26 & ---  & 1.2 \\ 
NGC4544 & 0.65 & 2.8 & 6.1 & ---  & 101.0 & 0.03 & ---  & 1.06 \\ 
VCC453 & 0.64 & 0.74 & 2.0 & ---  & 44.1 & 0.38 & ---  & 1.3 \\ 
NGC4353 & 0.62 & 6.31 & 4.7 & ---  & 79.1 & 0.0 & ---  & 0.88 \\ 
IC3094 & 0.58 & 1.07 & 2.3 & ---  & 92.5 & 0.19 & ---  & 3.49 \\ 
VCC1048 & 0.58 & 6.58 & 5.7 & ---  & 96.0 & 0.0 & ---  & 0.83 \\ 
NGC4571 & 0.56 & 0.7 & 11.3 & ---  & 209.2 & 0.42 & ---  & 1.23 \\ 
VCC297 & 0.55 & 6.53 & 3.9 & ---  & 67.5 & 0.0 & ---  & 0.7 \\ 
IC3259 & 0.55 & 6.39 & 5.7 & ---  & 89.8 & 0.0 & ---  & 0.65 \\ 
NGC4237 & 0.55 & 1.31 & 5.0 & ---  & 140.9 & 0.12 & ---  & 2.01 \\ 
VCC1725 & 0.54 & 1.24 & 3.8 & ---  & 47.7 & 0.14 & ---  & 0.35 \\ 
NGC4451 & 0.53 & 6.13 & 6.6 & ---  & 117.2 & 0.0 & ---  & 0.84 \\ 
    \hline
  \end{tabular}
\end{table*}

\begin{table*}
  \centering
  \caption{Table \ref{t:canpast1}, continued.}
  \label{t:canpast2} 
  \begin{tabular}{lcccccccc}
    \hline
 Name & $def$ & $d$ & $r_{\rm opt}$ & $r_{\rm HI, iso}$ & $v_{\rm rot}$ & $p_{\rm loc}$ & $p_{\rm max}$ & $p_{\rm def}$ \\  
    \hline \noalign{\rule{0pt}{0.6ex}}
VCC1529 & 0.52 & 2.63 & 2.9 & ---  & 32.0 & 0.03 & ---  & 0.23 \\ 
VCC568 & 0.51 & 6.54 & 3.3 & ---  & 85.0 & 0.0 & ---  & 1.24 \\ 
NGC4216 & 0.51 & 1.1 & 22.5 & 14.1 & 260.7 & 0.18 & 0.38 & 0.51 \\ 
NGC4416 & 0.5 & 1.36 & 5.4 & ---  & 186.2 & 0.12 & ---  & 2.66 \\ 
VCC1758 & 0.5 & 1.44 & 4.2 & ---  & 75.0 & 0.1 & ---  & 0.64 \\ 
IC3049 & 0.5 & 1.39 & 3.1 & ---  & 59.3 & 0.11 & ---  & 0.65 \\ 
VCC1789 & 0.5 & 2.3 & 2.7 & ---  & 54.0 & 0.04 & ---  & 0.68 \\ 
NGC4412 & 0.5 & 2.52 & 4.7 & ---  & 115.3 & 0.03 & ---  & 1.27 \\ 
VCC1596 & 0.49 & 1.0 & 0.9 & ---  & 30.8 & 0.21 & ---  & 1.41 \\ 
NGC4591 & 0.49 & 1.99 & 4.6 & ---  & 165.0 & 0.05 & ---  & 2.58 \\ 
NGC4502 & 0.49 & 1.28 & 3.7 & ---  & 104.3 & 0.13 & ---  & 1.47 \\ 
IC3066 & 0.48 & 1.17 & 2.5 & ---  & 67.6 & 0.16 & ---  & 1.18 \\ 
NGC4324 & 0.47 & 2.19 & 8.7 & ---  & 160.4 & 0.04 & ---  & 0.79 \\ 
VCC859 & 0.47 & 2.68 & 7.2 & ---  & 160.0 & 0.03 & ---  & 1.07 \\ 
NGC4694 & 0.46 & 3.75 & 6.2 & 3.4 & 34.7 & 0.01 & 0.07 & 0.06 \\ 
IC3471 & 0.46 & 1.08 & 2.1 & ---  & 60.8 & 0.18 & ---  & 1.13 \\ 
VCC168 & 0.45 & 1.18 & 1.1 & ---  & 50.6 & 0.15 & ---  & 2.28 \\ 
VCC324 & 0.44 & 2.67 & 3.3 & ---  & 56.1 & 0.03 & ---  & 0.42 \\ 
NGC4634 & 0.44 & 1.03 & 7.2 & ---  & 137.5 & 0.2 & ---  & 0.71 \\ 
VCC415 & 0.43 & 6.47 & 3.9 & ---  & 61.3 & 0.0 & ---  & 0.38 \\ 
NGC4430 & 0.43 & 6.49 & 10.1 & ---  & 150.9 & 0.0 & ---  & 0.47 \\ 
VCC1750 & 0.41 & 1.69 & 0.8 & ---  & 41.7 & 0.07 & ---  & 2.36 \\ 
NGC4466 & 0.4 & 1.4 & 3.0 & ---  & 88.6 & 0.11 & ---  & 1.11 \\ 
VCC468 & 0.4 & 2.56 & 1.4 & ---  & 47.4 & 0.03 & ---  & 1.11 \\ 
VCC130 & 0.4 & 1.39 & 1.6 & ---  & 63.0 & 0.11 & ---  & 1.61 \\ 
VCC1060 & 0.39 & 2.82 & 2.6 & ---  & 56.8 & 0.02 & ---  & 0.54 \\ 
NGC4343 & 0.39 & 6.42 & 8.3 & ---  & 173.3 & 0.0 & ---  & 0.75 \\ 
NGC4866 & 0.38 & 2.13 & 14.8 & ---  & 271.0 & 0.05 & ---  & 0.69 \\ 
IC3298 & 0.38 & 1.43 & 2.5 & ---  & 95.9 & 0.1 & ---  & 1.56 \\ 
VCC404 & 0.38 & 2.55 & 4.2 & ---  & 127.5 & 0.03 & ---  & 1.2 \\ 
IC3322 & 0.38 & 6.33 & 7.2 & ---  & 91.6 & 0.0 & ---  & 0.25 \\ 
VCC467 & 0.38 & 2.65 & 1.1 & ---  & 140.8 & 0.03 & ---  & 13.71 \\ 
VCC628 & 0.37 & 6.33 & 0.9 & ---  & 61.8 & 0.0 & ---  & 3.2 \\ 
NGC4533 & 0.37 & 3.0 & 6.4 & ---  & 86.5 & 0.02 & ---  & 0.26 \\ 
VCC1572 & 0.36 & 2.93 & 2.3 & ---  & 53.1 & 0.02 & ---  & 0.52 \\ 
VCC329 & 0.36 & 6.63 & 2.1 & ---  & 39.0 & 0.0 & ---  & 0.32 \\ 
VCC1918 & 0.35 & 2.15 & 2.5 & ---  & 35.5 & 0.04 & ---  & 0.19 \\ 
VCC888 & 0.35 & 6.23 & 3.9 & ---  & 41.2 & 0.0 & ---  & 0.13 \\ 
VCC1507 & 0.33 & 2.56 & 2.9 & ---  & 96.7 & 0.03 & ---  & 1.08 \\ 
NGC4630 & 0.33 & 2.64 & 5.7 & ---  & 101.9 & 0.03 & ---  & 0.38 \\ 
VCC772 & 0.32 & 2.4 & 1.3 & ---  & 33.0 & 0.03 & ---  & 0.46 \\ 
VCC693 & 0.31 & 2.2 & 2.9 & ---  & 95.7 & 0.04 & ---  & 0.98 \\ 
VCC741 & 0.31 & 2.61 & 2.1 & ---  & 72.6 & 0.03 & ---  & 0.95 \\ 
IC3099 & 0.3 & 0.99 & 4.6 & ---  & 105.5 & 0.22 & ---  & 0.53 \\ 
VCC1266 & 0.3 & 2.9 & 2.9 & ---  & 32.8 & 0.02 & ---  & 0.11 \\ 
NGC4212 & 0.3 & 1.18 & 8.9 & ---  & 152.3 & 0.15 & ---  & 0.37 \\ 
NGC4289 & 0.3 & 2.67 & 10.7 & ---  & 187.5 & 0.03 & ---  & 0.41 \\ 
VCC1685 & 0.3 & 2.79 & 5.3 & ---  & 64.5 & 0.03 & ---  & 0.15 \\ 
 \noalign{\rule{0pt}{0.6ex}}
VCC1011 & 0.29 & 1.43 & 3.2 & ---  & 71.5 & 0.1 & ---  & 0.43 \\ 
NGC4772 & 0.29 & 3.46 & 7.2 & 6.5 & 272.3 & 0.02 & 1.28 & 1.64 \\ 
IC3229 & 0.28 & 6.47 & 3.9 & ---  & 65.1 & 0.0 & ---  & 0.24 \\ 
IC3225 & 0.28 & 6.47 & 7.2 & ---  & 103.1 & 0.0 & ---  & 0.22 \\ 
IC3474 & 0.28 & 2.89 & 7.2 & ---  & 76.5 & 0.02 & ---  & 0.12 \\ 
NGC4378 & 0.27 & 2.25 & 7.6 & ---  & 284.3 & 0.04 & ---  & 1.52 \\ 
NGC4647 & 0.27 & 0.95 & 6.4 & ---  & 178.6 & 0.23 & ---  & 0.76 \\ 
NGC4207 & 0.26 & 1.39 & 4.8 & ---  & 108.9 & 0.11 & ---  & 0.44 \\ 
NGC4698 & 0.25 & 1.73 & 14.0 & 20.2 & 236.6 & 0.07 & 0.09 & 0.34 \\ 
VCC826 & 0.24 & 3.06 & 3.0 & ---  & 36.8 & 0.02 & ---  & 0.1 \\ 
IC3061 & 0.23 & 1.24 & 6.4 & ---  & 149.5 & 0.14 & ---  & 0.45 \\ 
VCC1257 & 0.23 & 1.49 & 3.4 & ---  & 79.1 & 0.1 & ---  & 0.36 \\ 
NGC4758 & 0.22 & 1.88 & 7.4 & ---  & 90.2 & 0.06 & ---  & 0.13 \\ 
VCC675 & 0.22 & 2.81 & 1.4 & ---  & 69.0 & 0.02 & ---  & 1.15 \\ 
VCC952 & 0.22 & 6.1 & 2.6 & ---  & 32.3 & 0.0 & ---  & 0.09 \\ 
VCC509 & 0.21 & 6.52 & 4.9 & ---  & 75.6 & 0.0 & ---  & 0.17 \\ 
    \hline
  \end{tabular}
\end{table*}

\begin{table*}
  \centering
  \caption{Table \ref{t:canpast1}, continued.}
  \label{t:canpast3} 
  \begin{tabular}{lcccccccc}
    \hline
 Name & $def$ & $d$ & $r_{\rm opt}$ & $r_{\rm HI, iso}$ & $v_{\rm rot}$ & $p_{\rm loc}$ & $p_{\rm max}$ & $p_{\rm def}$ \\  
    \hline \noalign{\rule{0pt}{0.6ex}}
VCC334 & 0.2 & 0.95 & 1.4 & ---  & 49.8 & 0.24 & ---  & 0.55 \\ 
NGC4390 & 0.19 & 6.07 & 7.3 & ---  & 114.8 & 0.0 & ---  & 0.18 \\ 
VCC1933 & 0.19 & 1.73 & 1.8 & ---  & 46.2 & 0.07 & ---  & 0.3 \\ 
VCC740 & 0.19 & 6.23 & 2.4 & ---  & 53.2 & 0.0 & ---  & 0.24 \\ 
NGC4779 & 0.17 & 1.86 & 5.2 & ---  & 217.7 & 0.06 & ---  & 1.04 \\ 
VCC1156 & 0.17 & 2.87 & 6.1 & ---  & 99.8 & 0.02 & ---  & 0.16 \\ 
VCC1468 & 0.16 & 2.32 & 2.5 & ---  & 52.5 & 0.04 & ---  & 0.19 \\ 
NGC4376 & 0.16 & 6.59 & 6.2 & ---  & 87.3 & 0.0 & ---  & 0.12 \\ 
NGC4316 & 0.16 & 6.18 & 8.3 & ---  & 157.5 & 0.0 & ---  & 0.24 \\ 
IC3268 & 0.14 & 6.47 & 6.5 & ---  & 58.9 & 0.0 & ---  & 0.04 \\ 
VCC459 & 0.13 & 1.7 & 2.1 & ---  & 68.9 & 0.07 & ---  & 0.38 \\ 
NGC4799 & 0.11 & 3.34 & 4.0 & ---  & 207.1 & 0.02 & ---  & 1.06 \\ 
NGC4423 & 0.09 & 6.55 & 10.2 & ---  & 85.0 & 0.0 & ---  & 0.03 \\ 
NGC4420 & 0.08 & 2.95 & 5.0 & ---  & 115.7 & 0.02 & ---  & 0.18 \\ 
VCC1141 & 0.07 & 6.12 & 1.5 & ---  & 69.1 & 0.0 & ---  & 0.44 \\ 
NGC4067 & 0.07 & 1.99 & 3.0 & ---  & 189.9 & 0.05 & ---  & 1.12 \\ 
VCC410 & 0.05 & 0.76 & 0.8 & ---  & 70.3 & 0.36 & ---  & 1.24 \\ 
VCC1873 & 0.05 & 1.9 & 1.4 & ---  & 40.5 & 0.06 & ---  & 0.15 \\ 
NGC4480 & 0.05 & 2.42 & 5.0 & ---  & 178.8 & 0.03 & ---  & 0.36 \\ 
VCC737 & 0.03 & 2.53 & 2.6 & ---  & 85.0 & 0.03 & ---  & 0.21 \\ 
VCC975 & 0.01 & 6.35 & 13.2 & ---  & 198.2 & 0.0 & ---  & 0.07 \\ 
    \hline
  \end{tabular}
\end{table*}

\begin{table*}
  \centering
  \caption{Like Table \ref{t:cannow}, but for non-deficient galaxies.}
  \label{t:cannodef} 
  \begin{tabular}{lcccccccc}
    \hline
 Name & $def$ & $d$ & $r_{\rm opt}$ & $r_{\rm HI, iso}$ & $v_{\rm rot}$ & $p_{\rm loc}$ & $p_{\rm max}$ & $p_{\rm def}$ \\  
    \hline \noalign{\rule{0pt}{0.6ex}}
VCC423 & -0.0 & 2.9 & 1.3 & ---  & 33.0 & 0.02 & ---  & ---  \\ 
NGC4635 & -0.02 & 2.39 & 4.5 & ---  & 110.3 & 0.04 & ---  & ---  \\ 
IC3617 & -0.05 & 1.46 & 3.2 & ---  & 52.5 & 0.1 & ---  & ---  \\ 
CGCG97067 & -0.05 & 3.88 & 6.5 & ---  & 107.6 & 0.01 & ---  & ---  \\ 
IC3881 & -0.05 & 2.64 & 9.9 & ---  & 104.6 & 0.03 & ---  & ---  \\ 
VCC827 & -0.06 & 6.37 & 12.0 & ---  & 140.5 & 0.0 & ---  & ---  \\ 
IC3576 & -0.06 & 1.76 & 5.3 & ---  & 43.1 & 0.07 & ---  & ---  \\ 
NGC4654 & -0.08 & 0.97 & 12.3 & 17.0 & 165.8 & 0.23 & 0.07 & ---  \\ 
NGC4746 & -0.08 & 1.53 & 5.4 & ---  & 170.8 & 0.09 & ---  & ---  \\ 
NGC4536 & -0.09 & 3.04 & 17.9 & 19.9 & 194.6 & 0.02 & 0.1 & ---  \\ 
IC3522 & -0.09 & 0.89 & 4.1 & ---  & 58.1 & 0.27 & ---  & ---  \\ 
IC3371 & -0.1 & 0.52 & 3.8 & ---  & 78.0 & 0.73 & ---  & ---  \\ 
NGC4303 & -0.11 & 2.44 & 16.3 & ---  & 131.7 & 0.03 & ---  & ---  \\ 
NGC4254 & -0.12 & 1.05 & 15.2 & 10.2 & 271.4 & 0.19 & 0.92 & ---  \\ 
NGC4178 & -0.13 & 1.39 & 13.2 & ---  & 132.3 & 0.11 & ---  & ---  \\ 
VCC848 & -0.15 & 6.57 & 3.9 & ---  & 137.3 & 0.0 & ---  & ---  \\ 
VCC1581 & -0.16 & 1.83 & 3.6 & ---  & 83.9 & 0.06 & ---  & ---  \\ 
NGC4123 & -0.18 & 3.28 & 12.4 & ---  & 165.4 & 0.02 & ---  & ---  \\ 
VCC1375 & -0.18 & 2.51 & 11.8 & ---  & 138.8 & 0.03 & ---  & ---  \\ 
NGC4186 & -0.19 & 1.44 & 2.3 & ---  & 75.0 & 0.1 & ---  & ---  \\ 
VCC566 & -0.2 & 6.27 & 2.4 & ---  & 49.5 & 0.0 & ---  & ---  \\ 
VCC1992 & -0.21 & 0.97 & 2.0 & ---  & 65.6 & 0.23 & ---  & ---  \\ 
NGC4519 & -0.22 & 1.13 & 8.9 & ---  & 119.7 & 0.17 & ---  & ---  \\ 
NGC4301 & -0.25 & 2.4 & 4.7 & ---  & 89.9 & 0.03 & ---  & ---  \\ 
IC3356 & -0.29 & 0.38 & 4.2 & ---  & 40.1 & 1.24 & ---  & ---  \\ 
NGC4116 & -0.3 & 3.34 & 9.4 & ---  & 137.1 & 0.02 & ---  & ---  \\ 
NGC4713 & -0.31 & 2.53 & 7.9 & 19.4 & 109.5 & 0.03 & 0.02 & ---  \\ 
NGC4527 & -0.34 & 2.9 & 14.5 & ---  & 193.8 & 0.02 & ---  & ---  \\ 
NGC4765 & -0.36 & 2.87 & 3.7 & ---  & 54.9 & 0.02 & ---  & ---  \\ 
VCC1091 & -0.37 & 6.18 & 4.9 & ---  & 90.8 & 0.0 & ---  & ---  \\ 
NGC4701 & -0.37 & 2.99 & 8.9 & ---  & 155.7 & 0.02 & ---  & ---  \\ 
NGC4532 & -0.38 & 1.78 & 6.4 & 12.1 & 85.4 & 0.07 & 0.04 & ---  \\ 
NGC4383 & -0.46 & 1.27 & 6.4 & 19.1 & 123.6 & 0.13 & 0.02 & ---  \\ 
NGC4561 & -0.49 & 2.09 & 3.7 & 12.5 & 164.9 & 0.05 & 0.06 & ---  \\ 
NGC4808 & -0.58 & 3.02 & 6.4 & 18.1 & 146.4 & 0.02 & 0.05 & ---  \\ 
    \hline
  \end{tabular}
\end{table*}

\bsp	
\label{lastpage}
\end{document}